\documentclass[PAPER, atlasdraft=false, cernpreprint=true, UKenglish, orcidlogo, texmf, orcidlogo]{atlasdoc}
\usepackage{atlaspackage}
\usepackage{atlasbiblatex}
\usepackage{multirow}
\usepackage{array,booktabs}
\newcolumntype{M}[1]{>{\centering\arraybackslash}m{#1}}
\usepackage{dcolumn}
\usepackage{atlasphysics}

\graphicspath{{logos/}{figures/}}

\usepackage{ANA-STDM-2020-08-PAPER-defs}

\AtlasTitle{Observation of \ensuremath{VVZ}\xspace production at $\sqrt{s}=13$ TeV with the ATLAS detector}

\AtlasAbstract{%
A search for the production of three massive vector bosons, $VVZ (V=W, Z)$, in proton--proton
collisions at $\sqrt{s} = 13$ TeV is performed using data with an integrated luminosity of $140$~fb$^{-1}$ recorded by the ATLAS detector at the Large Hadron Collider. Events produced in the leptonic final states $\ensuremath{WWZ}\xspace \to \ell\nu \ell\nu \ell \ell$ ($\ell=e, \mu$), $\ensuremath{WZZ}\xspace \to \ell\nu \ell\ell \ell\ell$, $\ensuremath{ZZZ}\xspace \to \ell\ell \ell\ell \ell\ell$, and the semileptonic final states $\ensuremath{WWZ}\xspace \to qq \ell\nu \ell \ell$ and $\ensuremath{WZZ}\xspace \to \ell\nu qq \ell \ell$, are analysed. The measured cross section for the $pp \rightarrow VVZ$ process is $660^{+93}_{-90}(\text{stat.})^{+88}_{-81}(\text{syst.})$ fb, and the observed (expected) significance is 6.4 (4.7) standard deviations, representing the observation of $VVZ$ production. In addition, the measured cross section for the $pp \rightarrow WWZ$ process is $442 \pm 94 (\text{stat.})^{+60}_{-52}(\text{syst.})$ fb, and the observed (expected) significance is 4.4 (3.6) standard deviations, representing evidence of $WWZ$ production. The measured cross sections are consistent with the Standard Model predictions. Constraints on physics beyond the Standard Model are also derived in the effective field theory framework by setting limits on Wilson coefficients for dimension-8 operators describing anomalous quartic gauge boson couplings.}

\AtlasRefCode{STDM-2020-08}

\PreprintIdNumber{CERN-EP-2024-321}

\AtlasJournalRef{\PLB 866 (2025) 139527}
\AtlasDOI{10.1016/j.physletb.2025.139527}
\hypersetup{pdftitle={ATLAS document},pdfauthor={The ATLAS Collaboration}}

\begin{document}

\setcounter{tocdepth}{4}
\setcounter{secnumdepth}{4}

\maketitle

\section{Introduction}
In the Standard Model (SM), the non-Abelian structure of the electroweak sector predicts the self-interaction of vector bosons leading to vertices with three or four vector bosons. Studying processes containing vertices with four vector bosons is a sensitive test of the SM as deviations from the SM expectation would be hints of new physics at higher energy scales~\cite{Belyaev:1998ih, Du:2012vh, Fichet:2013ola, Giudice:2007fh}. Four vector boson vertices can be probed in vector boson scattering processes~\cite{STDM-2017-06,CMS-SMP-19-012} or in three boson final states.

The production of three massive vector bosons in proton--proton ($pp$) collisions at the Large Hadron Collider (LHC)~\cite{Evans:2008zzb} was studied using data collected at a centre-of-mass energy of $\sqrt{s}= \SI{13}{\TeV}$ by both the ATLAS~\cite{PERF-2007-01} and CMS~\cite{CMS-CMS-00-001} experiments.
ATLAS has provided evidence for the production of $WWW$ and $WVZ$ ($V=W, Z$) using data with an integrated luminosity of 79.8 fb$^{-1}$~\cite{STDM-2017-22}, and later observed $WWW$ production using 139 fb$^{-1}$ of data~\cite{STDM-2019-09}. Meanwhile, CMS has observed the combined production of $VVV$ ($WWW+WWZ+WZZ+ZZZ$) based on 137 fb$^{-1}$ of data~\cite{CMS-SMP-19-014}. This article reports on a search for the production of three massive vector bosons of which at least one is a \Z boson, i.e. \WWZs, \WZZs, and \ZZZ, using 140 fb$^{-1}$ of data.

\begin{figure}[b]
\centering
\subfloat[ ]{\includegraphics[width=0.20\textwidth]{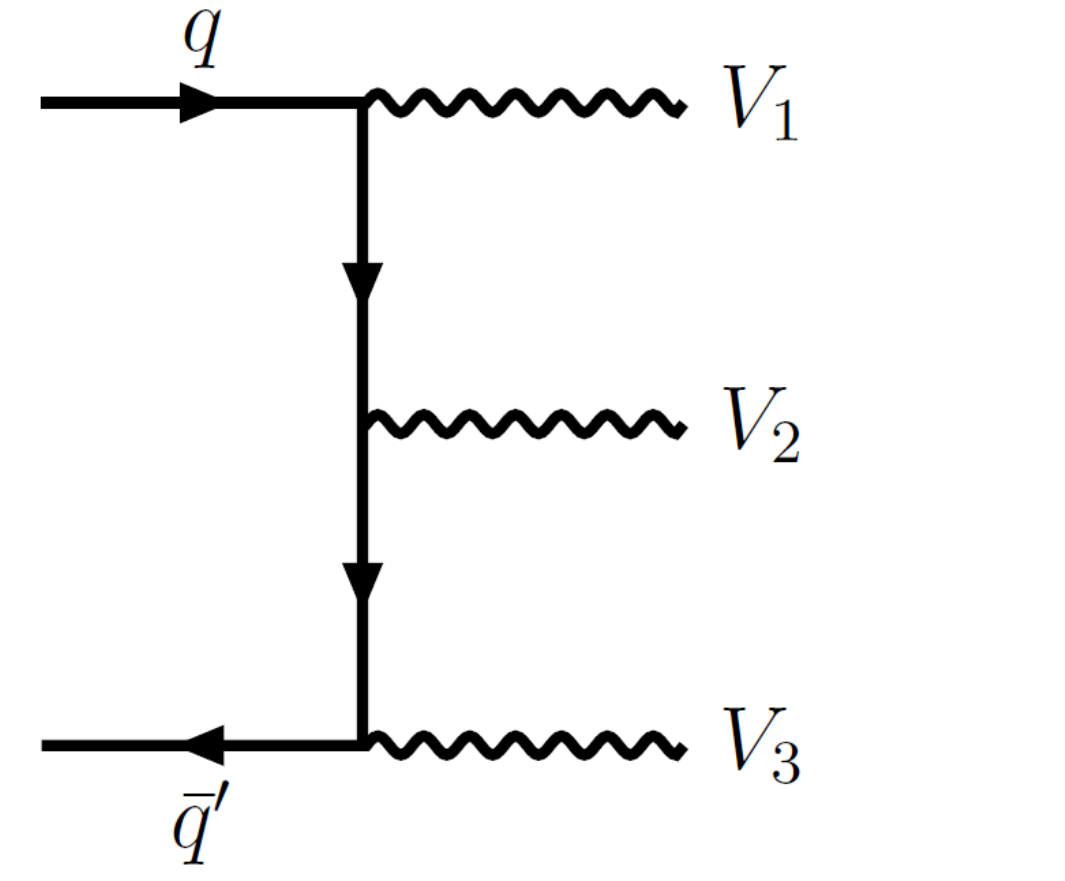}}  \hspace{0.05em}
\subfloat[ ]{\includegraphics[width=0.23\textwidth]{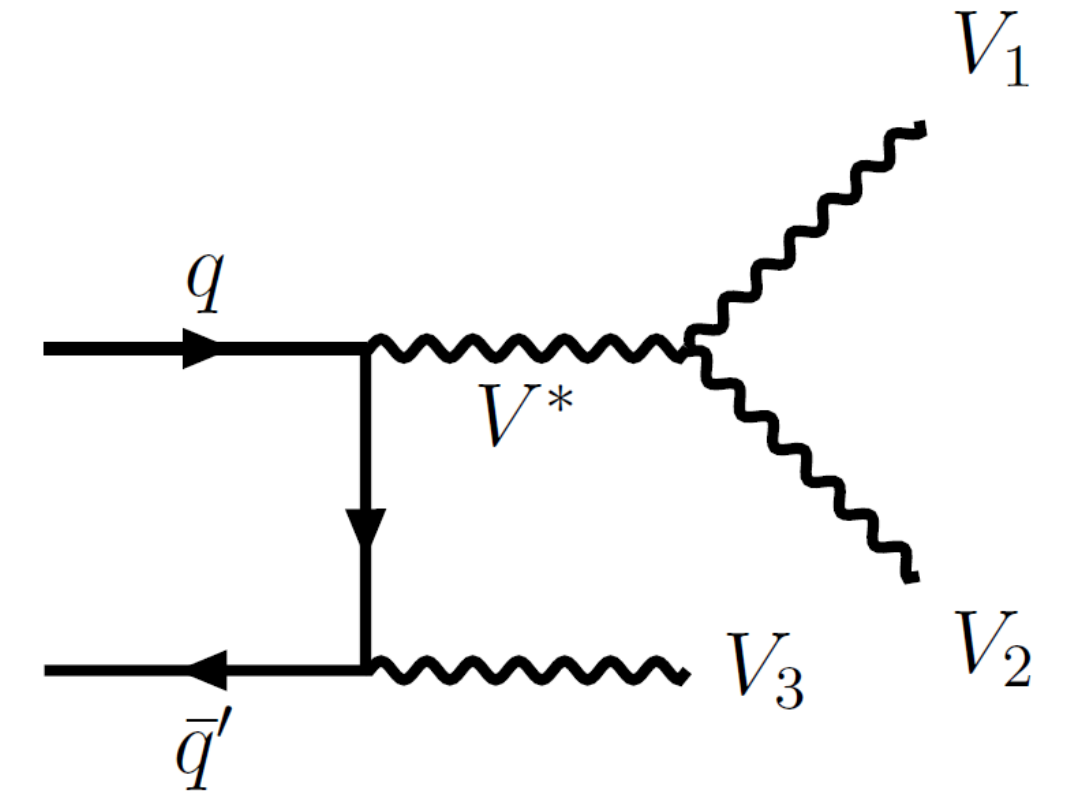}}  \hspace{0.5em}
\subfloat[ ]{\includegraphics[width=0.23\textwidth]{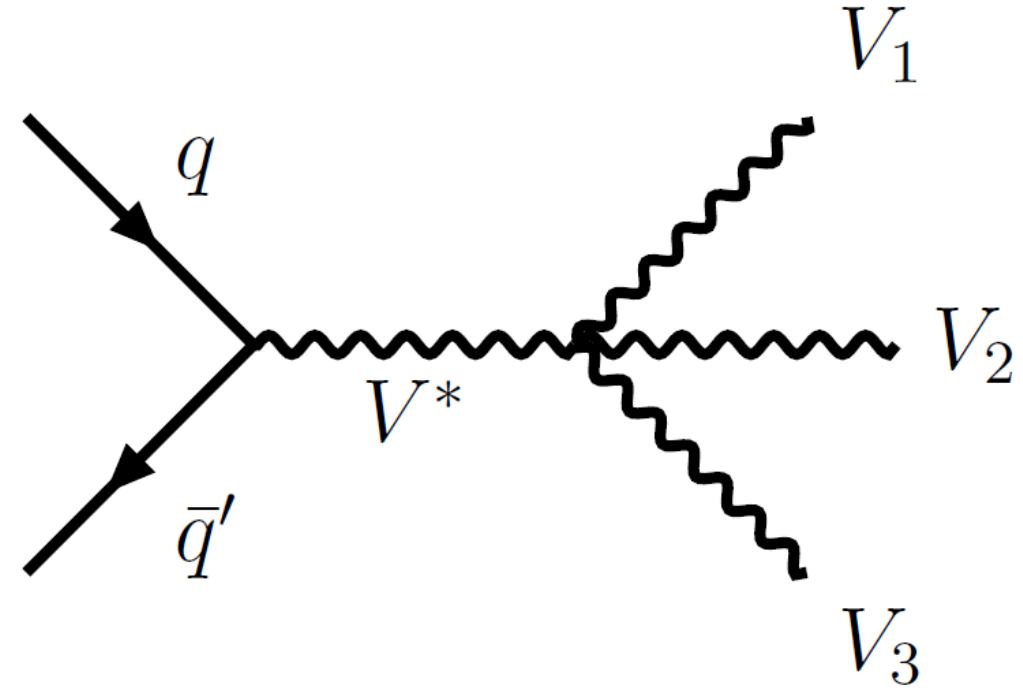}}  \hspace{0.5em}
\subfloat[ ]{\includegraphics[width=0.23\textwidth]{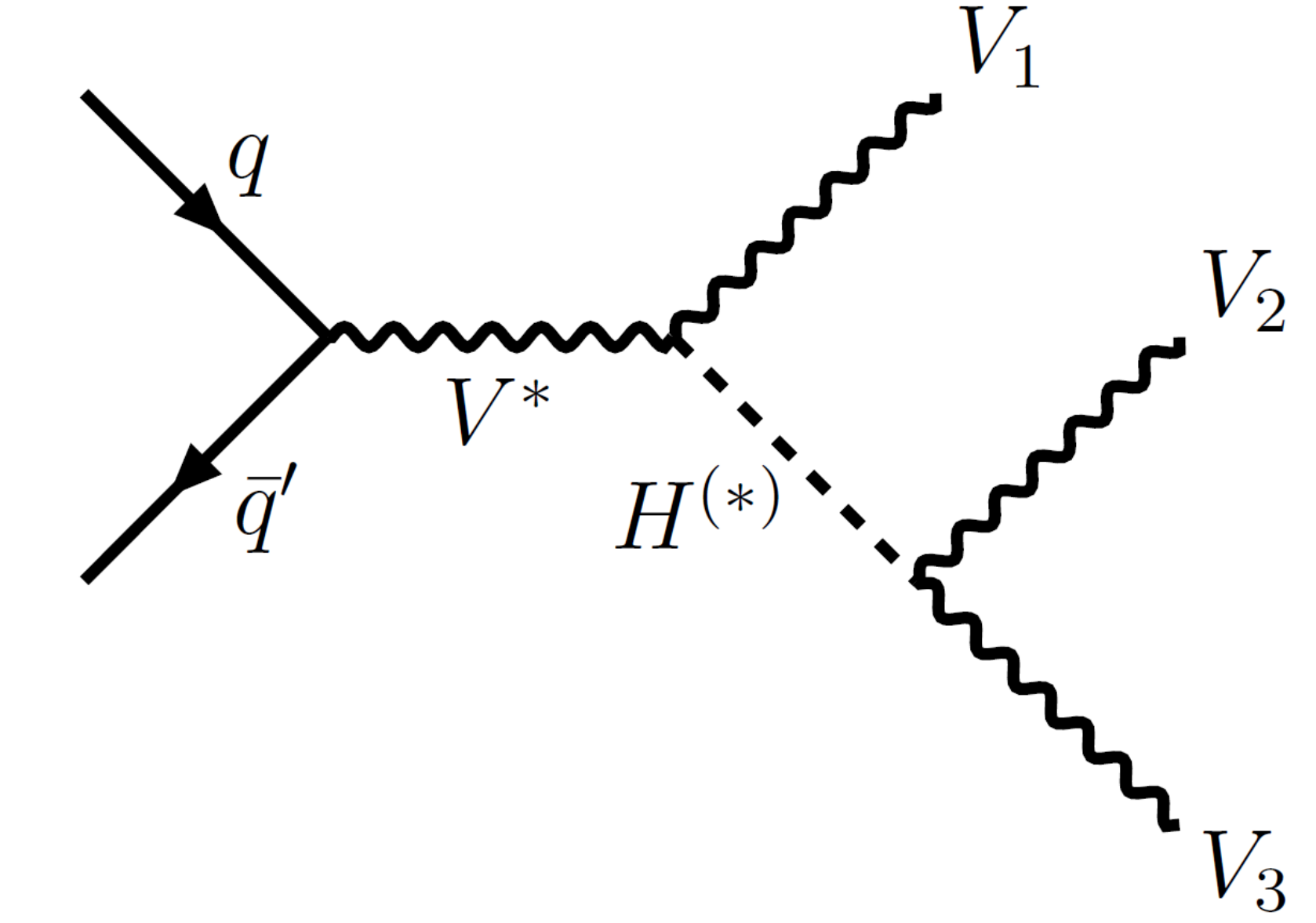}}
\caption{Representative Feynman diagrams for the production of three massive vector bosons, including diagrams with (a) mono boson vertices, diagrams sensitive to (b) triple and (c) quartic gauge boson couplings, and (d) the Higgsstrahlung process, where $H^{(*)}$ denotes an on-shell or off-shell Higgs boson.}
\label{fig:feyn}
\end{figure}

The production of three massive vector bosons can occur via diagrams containing mono to quartic boson interaction vertices, and via the Higgsstrahlung process. Representative Feynman diagrams are shown in Figure~\ref{fig:feyn}.

Three distinct search channels based on the number of leptons in the final state are used, targeting different tri-boson final states, namely \threeLeptonJets for \WWZs and \WZZs, \fourlepton for \WWZs, and at least \fivelepton for \WZZs and \ZZZ processes, with $\ell=e,\mu$. The sets of selection criteria for the different search channels are designed to avoid overlap between the channels.
The dominant background processes are \WZs production for the \threelepton channel, and \WZs and \ZZ~production for the \fourlepton and \fivelepton channels. To enhance the sensitivity, the signal regions for the \threelepton and \fourlepton channels are further split into sub-categories.
Boosted decision tree (BDT) discriminants are trained individually for each channel to enhance the separation between signal and background events.
These channels are then combined with a binned maximum-likelihood fit of the BDT discriminants, yielding a combined signal strength parameter $\mu$ for \VVZ production with $\mu$ defined as the ratio of the measured tri-boson production cross section to its SM prediction. The signal strength is also determined separately for \WWZs and \WZZ production. Additionally, the signal strength for on-shell \VVZ production is extracted by treating the Higgsstrahlung contribution as background.
Finally, limits on effective field theory (EFT) parameters describing the four-vector-boson interaction vertex are derived with discriminants optimised to enhance the sensitivity to EFT contributions to the SM processes.

\section{The ATLAS detector, data and simulation samples }

The ATLAS detector at the LHC is a multipurpose particle detector
with a forward--backward symmetric cylindrical geometry and a near \(4\pi\) coverage in
solid angle\footnote{ATLAS uses a right-handed coordinate system with its origin at the nominal interaction point (IP)
in the centre of the detector and the \(z\)-axis along the beam pipe.
The \(x\)-axis points from the IP to the centre of the LHC ring, and the \(y\)-axis points upwards.
Polar coordinates \((r,\phi)\) are used in the transverse plane,
\(\phi\) being the azimuthal angle around the \(z\)-axis.
The pseudorapidity is defined in terms of the polar angle \(\theta\) as \(\eta = -\ln \tan(\theta/2)\).
Transverse momentum (\pT) is defined relative to the beam axis and is calculated as $\pT = p\sin\theta$ where $p$ is the momentum.
Transverse energy (\ET) is calculated as $\ET = E\sin\theta$ where $E$ is the energy.
Angular distance is measured in units of \(\Delta R \equiv \sqrt{(\Delta\eta)^{2} + (\Delta\phi)^{2}}\).}.
It consists of an inner tracking detector (ID) surrounded by a thin superconducting solenoid
providing a 2~T axial magnetic field, electromagnetic (EM) and hadron calorimeters, and a muon spectrometer (MS).
The inner tracking detector covers the pseudorapidity range of \(|\eta| < 2.5\).
It consists of silicon pixel, silicon microstrip, and transition radiation tracking detectors.
Lead/liquid-argon (LAr) sampling calorimeters provide EM energy measurements
with high granularity. A steel/scintillator-tile hadron calorimeter covers the central pseudorapidity range \(|\eta| < 1.7\).
The endcap and forward regions are instrumented with LAr calorimeters
for both the EM and hadronic energy measurements up to \(|\eta| = 4.9\).
The muon spectrometer surrounds the calorimeters and is based on
three large superconducting air-core toroidal magnets with eight coils each.
The field integral of the toroids ranges between \num{2.0} and $6.0\,\mathrm{Tm}$
across most of the detector.
The muon spectrometer includes a system of precision tracking chambers and fast detectors for triggering.
A two-level trigger system~\cite{TRIG-2016-01} is used to select interesting events.
The first-level trigger is implemented in hardware and uses a subset of the detector information
to accept events at a rate below $100\,\mathrm{kHz}$.
This is followed by a software-based trigger that
reduces the accepted event rate to $1\,\mathrm{kHz}$ on average,
depending on the data-taking conditions.
An extensive software suite~\cite{SOFT-2022-02} is used in data simulation, in the reconstruction
and analysis of real and simulated data, in detector operations, and in the trigger and data acquisition
systems of the experiment.

The data used in the analysis were collected between 2015 and 2018 in $pp$ collisions at
$\sqrt{s} = \SI{13}{\TeV}$.  Only events recorded with a fully operational
detector and stable beams are included resulting in a total integrated luminosity of \lumi~\cite{DAPR-2021-01}.
Candidate events are selected by single or multiple lepton ($e$ or $\mu$) triggers with transverse momentum thresholds varying depending on the lepton flavour, isolation requirements, and run period. Due to the presence of at least three leptons in the final state and the requirement of at least one lepton with $\pT > \SI{27}{\GeV}$ in the event selection, the lepton triggers are fully efficient for the tri-boson signals
in the signal regions defined in Section~\ref{sec:event_selection}.

Signal and background processes were simulated with a range of Monte Carlo (MC)
event generators and the ATLAS detector response~\cite{SOFT-2010-01}
was modelled with \GEANT~\cite{Agostinelli:2002hh}.  The effect
of multiple $pp$ interactions in the same and neighbouring bunch crossings
(pile-up) was included by overlaying minimum-bias events simulated with
\PYTHIA{8.186}~\cite{Sjostrand:2007gs}
using the \tune{A3}~\cite{ATL-PHYS-PUB-2016-017} set of tuned MC parameters and the \pdf{NNPDF2.3LO}~\cite{Ball:2012cx} parton distribution function (PDF) set, on each generated event in all samples.
Tri-boson signal and \WWW background events~\cite{ATL-PHYS-PUB-2017-005} with three
on-mass-shell vector bosons including processes involving an off-shell Higgs boson mediator were simulated using \SHERPA{2.2.2}~\cite{Bothmann:2019yzt} with the \pdf{NNPDF3.0NNLO}~\cite{Ball:2014uwa} PDF.  Off-mass-shell tri-boson final states with an on-shell
Higgs boson mediator, i.e. $WH \rightarrow WVV^*$ and $ZH \rightarrow ZVV^*$, were generated
using \POWHEGBOX{2}~\cite{powheg, Nason:2004rx, Frixione:2007vw,
Cullen:2011ac, Hamilton:2012np, Luisoni:2013kna} interfaced to \PYTHIA{8.186} and \EVTGEN{1.6.0}~\cite{evtgen}.
All tri-boson processes were generated at next-to-leading-order
(NLO) quantum chromodynamics (QCD) accuracy~\cite{Lazopoulos:2007ix, Binoth:2008kt, Hankele:2007sb,Campanario:2008yg}.
The expected cross section ($VH$ included) is \xsecwwz\ for \WWZ, \xsecwzz\ for \WZZ, and \xseczzz\ for \ZZZ.
The total theory uncertainty in the signal cross sections is about 10\% and was evaluated by varying parameters in the simulation related to the
renormalisation and factorisation scales, parton shower and PDF sets.
For the interpretation within the EFT approach (described in Section~\ref{s:eft}), samples of \WWZ and \WZZ production with
non-zero Wilson coefficients for the $M2$, $M3$, $M4$ and $M5$ operators
in the Eboli model~\cite{PhysRevD.74.073005} were generated at leading-order (LO) with \MADGRAPH{2.7.3}~\cite{Alwall:2014hca} using the \pdf{NNPDF3.0NLO} PDF set and interfaced to \PYTHIA{8.245}~\cite{Sjostrand:2014zea} and the \tune{A14}~\cite{ATL-PHYS-PUB-2014-021} tune. The EFT interpretation does not include the $5\ell$ channel, so the \ZZZ samples are not included as their impact is negligible in the $3\ell$ and $4\ell$ channels.

Diboson ($WW$, $WZ$, $ZZ$, $W\gamma$, and $Z\gamma$) processes~\cite{ATL-PHYS-PUB-2017-005} were modelled
using \SHERPA{2.2.12} with the \pdf{NNPDF3.0NNLO} PDF set.  Single boson ($W/Z$+jets)~\cite{ATL-PHYS-PUB-2016-003}
production and electroweak production of $W^\pm W^\pm+2$ jets, $WZ+2$ jets, and $ZZ+2$ jets, were modelled
using \SHERPA{2.2.11} and \SHERPA{2.2.2}, respectively.

Top-quark pair events (\ttbar) were simulated using \POWHEGBOX{2}~\cite{Frixione:2007nw} interfaced to
\PYTHIA{8.230} and \EVTGEN{1.6.0}. The \pdf{NNPDF3.0NLO} PDF set was used for the matrix-element calculation,
while the \pdf{NNPDF2.3LO} PDF set was used for the showering with the \tune{A14} tune. Production processes of
a top-quark pair in association with a vector boson (\ttZ and \ttW) were modelled at NLO with \MADGRAPH{2.2.3}
with the \pdf{NNPDF3.0NLO} PDF set and interfaced to \PYTHIA{8.210} using the \tune{A14} tune and \EVTGEN{1.2.0} for heavy flavour decays.
The \ttH production was generated at LO in QCD with \POWHEGBOX{2} interface to \PYTHIA{8.210} and \EVTGEN{1.6.0}.
Other background processes containing top quarks were generated with \MGNLO~\cite{amcnlo} interfaced to \PYTHIA{8.230}, at LO
($\ttbar\gamma$, $tZ$, $t\bar{t}WW$, $t\bar{t}WZ$, and $t\bar{t}t\bar{t}$) or with \POWHEGBOX{2}~\cite{Re:2010bp} interfaced to \PYTHIA{8.212} and \EVTGEN{1.6.0} ($tWZ$).

\section{Object definitions, event selection and background estimation}
\subsection{Object definitions and preselection criteria}

At the trigger level, candidate events are selected using all combinations of unprescaled single lepton, dilepton, tri-lepton, and four-lepton triggers~\cite{TRIG-2018-01, TRIG-2018-05}.
Triggers requiring a mixture of lepton flavours (for example, dilepton triggers with an electron and a muon at the trigger level) are also used.
These triggers require leptons to satisfy certain $p_T$ threshold, identification, and isolation criteria.
The combined trigger efficiency is fully efficient for tri-boson events in the fiducial regions.
The candidate events are required to contain one reconstructed primary vertex~\cite{PERF-2015-01}. If more than one reconstructed vertex is found, the vertex with the largest $\pT^2$ sum of associated ID tracks is considered as the primary vertex.

Electrons are reconstructed from energy clusters in the EM calorimeter matched to ID tracks and are identified using a likelihood discriminant constructed with information about the shape of the EM showers in the calorimeter, the track properties, and the quality of the track-to-cluster matching for the candidate~\cite{EGAM-2018-01}. Electrons must satisfy the ``Loose'' requirement based on the electron likelihood discriminant and have at least one hit in the innermost pixel layer.
Electrons are required to have $\pT>7$ \GeV and $|\eta|<2.47$. Electron candidates reconstructed within the transition region between the barrel and endcap calorimeters, $1.37 < |\eta| < 1.52$, are also kept.

Muons are reconstructed in multiple ways based on information from the ID, the MS, and the calorimeters~\cite{MUON-2018-03}.
In the range of the ID coverage, the muon reconstruction is primarily performed by a global fit of fully reconstructed tracks in the ID and the MS (referred to as ``combined muons''). In the central region ($|\eta| < 0.1$) of the detector where the MS lacks in coverage, muons can also be identified by matching a fully reconstructed ID track to either an MS track segment (referred to as ``segment-tagged muons'') or a calorimetric energy deposit consistent with that of a minimum-ionizing particle (referred to as ``calorimeter-tagged muons''). For the last two cases, the muon momentum is determined by the ID track alone. In the forward MS region ($2.5 < |\eta| < 2.7$), MS tracks with hits in the three MS layers are accepted (referred to as ``standalone muons''). Muons are required to satisfy the ``Loose'' identification requirement described in Ref.~\cite{MUON-2018-03} and to have $|\eta|<2.7$ and $\pT>5$ \GeV for all types except for the calorimeter-tagged type where the requirement is increased to $\pT>15$ \GeV.

Both electrons and muons are required to be consistent with originating from the primary vertex by imposing requirements on the transverse impact parameter, $d_0$, its uncertainty, $\sigma_{d_0}$, the longitudinal impact parameter, $z_0$, and the polar angle $\theta$.  These requirements are $|d_{0}|/\sigma_{d_0}<5$ and $|z_{0} \times \sin\theta| < \SI{0.5}{mm}$ for electrons, and $|d_{0}|/\sigma_{d_0}<3$ and $|z_{0} \times \sin\theta| < \SI{0.5}{mm}$ for muons. The leptons selected with these criteria are classified as baseline leptons.

Jets are reconstructed from particle-flow objects using the anti-$k_t$ algorithm~\cite{Cacciari:2008gp, Cacciari:2011ma} with a radius parameter of 0.4.
The particle-flow algorithm combines information about ID tracks and energy deposits in the calorimeters to form the input for jet reconstruction~\cite{PERF-2015-09}. Jets are selected by requiring $\pT > 20$ \GeV and $|\eta| < 4.5$. To suppress jets arising from pile-up, a jet-vertex-tagging (JVT) technique~\cite{PERF-2014-03} using a
multivariate likelihood is applied to jets with $\pT < 60$ \GeV and $|\eta| < 2.4$, ensuring that the selected
jets are matched to the primary vertex.

Jets containing $b$-hadrons are identified ($b$-tagged) using the DL1r $b$-tagging algorithm~\cite{FTAG-2019-07} based on a deep neural network that combines information from displaced tracks and reconstructed secondary and tertiary vertices inside jets. A jet with $|\eta|<2.5$ is $b$-tagged if the response value of the DL1r algorithm exceeds a predefined threshold. The operating point defined with an efficiency of 77\% for $b$-jets measured in simulated $t\bar{t}$ events is used.

The missing transverse momentum, whose magnitude is denoted \met, is defined as
the negative vector sum of the \pt of all reconstructed and calibrated physics objects in the event~\cite{JETM-2020-03}.
This sum includes a term to account for the energy from
low-momentum particles that are not associated with a reconstructed lepton or jet.

The object reconstruction and identification algorithms do not always result in
unambiguous identifications of physics objects. An overlap removal procedure is applied to the leptons and jets in the following order.
Calorimeter-tagged muons sharing a track with any electrons are removed.
Electrons sharing an ID track with muons are removed. Any jet within $\Delta R
< 0.2$ of an electron is removed and electrons within $\Delta R < 0.4$ of any
remaining jets are removed.  Jets with less than three associated tracks and
within $\Delta R < 0.2$ of a muon are removed, and muons within $\Delta R <0.4$ of any of the remaining jets are removed.

The preselected events are required to have at least three charged leptons satisfying the baseline lepton selection criteria.
In addition, at least one same-flavour opposite-sign charge (SFOS) lepton pair that is consistent with the $Z$ boson pole mass ($m_Z=91.188$ \GeV) within 40 \GeV is required. Table~\ref{table:pre-selection} shows all preselection criteria used.

\begin{table}[t]
\caption{Summary of the object selection and event preselection criteria used in the analysis.}
\begin{center}
\resizebox{0.8\textwidth}{!}{
\begin{tabular}{c|c}
\hline \hline
\multicolumn{2}{c}{\textbf{Object selection criteria}} \\ \hline
\multirow{3}{*}{Electron}      & Passes the ``Loose'' quality requirement and have at least one hit in the innermost pixel layer\\
& $|d_{0}|/\sigma_{d_0}<5$, $|z_{0} \times \sin\theta| < \SI{0.5}{mm}$ \\
& $\pT>7$ \GeV, $|\eta|<2.47$ \\ \hline
\multirow{3}{*}{Muon}          & Passes the ``Loose'' quality requirement \\
& $|d_{0}|/\sigma_{d_0}<3$, $|z_{0} \times \sin\theta| < \SI{0.5}{mm}$ \\
&  $\pT>5$ \GeV ($\pT>15$ \GeV for calorimetr-tagged muons), $|\eta|<2.7$  \\ \hline
Jet                            & Passes the JVT requirement, $\pT>20$ \GeV, $|\eta|<4.5$ \\ \hline \hline
\multicolumn{2}{c}{\textbf{Event preselection criteria}} \\ \hline
Trigger                        & Single lepton, dilepton, tri-lepton, or quad-lepton triggers   \\   \hline
Number of charged leptons      & $\ge 3$                      \\ \hline
$Z$ boson invariant mass      & $|m_{\ell\ell} - m_Z| < 40$~\GeV               \\
\hline \hline
\end{tabular}
}
\end{center}
\label{table:pre-selection}
\end{table}

\subsection{Event selection criteria}
\label{sec:event_selection}
The \threelepton channel requires exactly three charged leptons and at least one reconstructed jet in the final state. All SFOS dilepton pairs should have an invariant mass $m_{\ell\ell}>12$ \GeV. Events are required to have at least one SFOS lepton pair that is consistent with $m_Z$ within 20 \GeV. This pair is considered to be the $Z$ boson candidate. If more than one pair can be formed, the pair whose invariant mass is closest to $m_Z$ is taken as the $Z$ boson candidate. These two leptons are assigned to the $Z$ boson decay (labelled $Z$-leptons). The remaining third lepton is assigned to the $W$ boson decay (labelled $W$-lepton). To reduce instrumental backgrounds, events with both $Z$-electrons falling into the calorimeter transition region $1.37<|\eta|<1.52$ are rejected. In addition, events with the third electron falling into the calorimeter transition region are rejected.

All three leptons are required to have $\pT>15$ \GeV and at least one must have $\pT>27$ \GeV. The electrons must satisfy the ``Loose'' isolation requirement based on prompt tracks within a shrinking cone around the electron track as $p_T$ increases and calorimeter energy depositions within $\Delta R=0.2$ of the electron. Similarly, the muons are required to satisfy the ``Loose'' isolation requirement based on prompt tracks and calorimeter energy depositions~\cite{MUON-2018-03}.
Backgrounds originating from misidentified leptons are suppressed by requiring the $W$-lepton to satisfy more stringent selection criteria. The $W$-lepton must satisfy the ``Tight'' quality requirement for electrons based on thelikelihood discriminant~\cite{PERF-2017-01} and the ``Tight'' quality requirement for muons~\cite{MUON-2018-03}. Both $W$-leptons must satisfy the ``Tight'' isolation working point based on a multivariate discriminant used to reject non-prompt leptons from heavy-flavour decays~\cite{HIGG-2017-02}. At least one jet with $\pT>20$ \GeV and $|\eta|<2.5$ is required to be present in the final state.
Events with at least one $b$-jet are rejected.

To gain sensitivity, the \threelepton channel makes use of events containing four leptons that satisfy the \threelepton channel selection shown in Table~\ref{table:EventSelection_3l}, but fail to satisfy the \fourlepton channel selection criteria shown in Table~\ref{table:EventSelection_4l}. These \fourlepton events are required to contain only one lepton pair with $|m_{\ell\ell}-m_Z|<20$ \GeV. %
One of the other two leptons is required to satisfy the \threelepton quality requirements for a $W$-lepton, while the other lepton is required to fail this requirement. This increases the signal yield by 2\% and the estimated background by 1.5\%.

Three signal regions (SRs) are defined according to the number of jets and the invariant mass of the two leading jets in the event:
the \TLjone SR is defined by requiring events with exactly one reconstructed jet; the requirements of at least two jets with the invariant mass of the two leading jets, $m_{jj}$, lying within a window close to the $W$ or $Z$ boson pole mass, i.e. $60<m_{jj}<110$ \GeV, defines the \TLjtwoin SR; the remaining events compose the \TLjtwoout SR.
With these selection criteria applied, the signal-to-background ratio is close to 1\% for the \TLjone channel, 5\% for the \TLjtwoin channel, and 3\% for the \TLjtwoout channel. To reduce the $WZ+$jets background, each event is required to have a BDT score (described in Section~\ref{sec:results}) that is larger than 0.42. Table~\ref{table:EventSelection_3l} summarises all selection criteria used and the definitions of these three SRs.

\begin{table}[t]
\caption{Overview of the criteria used to select inclusive $3\ell$ events and the three \threelepton SRs.}
\begin{center}
\resizebox{0.95\textwidth}{!}{
\begin{tabular}{c|c|c|c}
\hline \hline
\multicolumn{4}{c}{\textbf{Inclusive $3\ell$ event selection}} \\
\hline
Satisfy preselection criteria   & \multicolumn{3}{c}{\checkmark}   \\
Lepton      & \multicolumn{3}{c}{$\pT>15$ \GeV and at least one lepton with $\pT>27$ \GeV}            \\
Lepton from the $Z$ decays  & \multicolumn{3}{c}{``Loose'' isolation requirement based on tracker and calorimeter measurements}  \\
Lepton from the $W$ decays       & \multicolumn{3}{c}{``Tight'' identification and ``Tight'' isolation}                \\
Invariant mass of any SFOS dilepton pairs  & \multicolumn{3}{c}{$>12$ \GeV} \\
Invariant mass of the $Z$ boson    & \multicolumn{3}{c}{$|m_{\ell\ell} - m_Z|<20$~\GeV}               \\
Number of leptons & \multicolumn{3}{c}{$=3$} \\
Number of $b$-jets       & \multicolumn{3}{c}{$=0$}      \\

\bottomrule
\multicolumn{4}{c}{\textbf{$3\ell$ signal regions}} \\[0.5ex]
\hline
& \textbf{\TLjone} & \textbf{\TLjtwoin} & \textbf{\TLjtwoout} \\[0.5ex]
\hline
Satisfy inclusive $3\ell$ selection criteria & \checkmark &\checkmark & \checkmark \\
BDT score $>0.42$ & \checkmark &\checkmark & \checkmark \\
Number of jets &  $=1$ & $\ge 2$ & $\ge 2$   \\
$m_{jj}$ & {$~~~~~~-~~~~~~$} & {$>60$ \GeV and $<110$~\GeV}  & {$<60$~\GeV or $>110$ \GeV} \\
\hline \hline
\end{tabular}
}
\end{center}
\label{table:EventSelection_3l}
\end{table}

The \fourlepton channel requires exactly four leptons in the event. The SFOS lepton pair with its invariant mass, \mll, closest to $m_Z$ is identified as the \Z boson candidate and must fulfil $|\mll - m_Z| < 20$ \GeV. The remaining two leptons are assigned as leptons from $W$ decays and must satisfy the ``Tight'' isolation and ``Medium''
quality requirements. All SFOS lepton pairs must have an invariant mass of $\mll > 12$ \GeV. The set of four leptons is required to satisfy the ordered \pt thresholds of 30, 15, 8, and 6 \GeV. The angular distance between any two leptons is required to be $\dR > 0.1$.  Furthermore, each event must have $\MET > 10$ \GeV and no jets reconstructed as $b$-tagged jets.
Events that satisfy the selection are categorised into three SRs depending on the flavours of the two $W$-leptons. Those events where the two leptons have different flavour compose the \FLDF SR and events with two same-flavour leptons are split into two SRs based on their invariant mass: events satisfying $|m_{\ell\ell}-m_Z|<20$ \GeV define the \FLSFZ SR, while the remaining events compose the \FLSFnoZ SR.
Table~\ref{table:EventSelection_4l} summarises the selection criteria and the definitions of these three SRs.

\begin{table}[t]
\caption{Overview of the criteria used to select inclusive $4\ell$ events and the three $4\ell$ SRs.}
\begin{center}
\resizebox{0.99\textwidth}{!}{
\begin{tabular}{c|c|c|c}
\hline \hline
\multicolumn{4}{c}{\textbf{Inclusive $4\ell$ event selection}} \\
\hline
Satisfy preselection criteria   & \multicolumn{3}{c}{\checkmark}   \\
Lepton      & \multicolumn{3}{c}{Exactly four leptons with $\pT>30,15,8,6$ \GeV}            \\
Lepton from the $Z$ decays  & \multicolumn{3}{c}{``Loose'' isolation requirement based on tracker and calorimeter measurements}  \\
Leptons from the $W$ decays    & \multicolumn{3}{c}{``Medium'' identification and ``Tight'' isolation}                \\
Invariant mass of any SFOS dilepton pairs  & \multicolumn{3}{c}{$>12$~\GeV} \\
Invariant mass of the $Z$ boson     & \multicolumn{3}{c}{$|m_{\ell\ell} - m_Z|<20$~\GeV}               \\
Minimum angular distance between any lepton pairs  & \multicolumn{3}{c}{$>0.1$} \\
\met  & \multicolumn{3}{c}{$>10$ \GeV} \\
Number of $b$-jets       & \multicolumn{3}{c}{$=0$}                                                 \\
\bottomrule
\multicolumn{4}{c}{\textbf{$4\ell$ signal regions}} \\[0.5ex]
\hline
& \textbf{\FLDF} & \textbf{\FLSFZ} & \textbf{\FLSFnoZ} \\[0.5ex]
\hline
Satisfy inclusive $4\ell$ selection criteria & \checkmark &\checkmark & \checkmark \\
Flavour for lepton from the $W$ decays & $e \mu$  & same-flavour & same-flavour   \\
\multirow{1}{*}{$m_{\ell\ell}$ for the two $W-$leptons} & \multirow{1}{9em}{$~~~~~~~~~~~~~~~~-~$} & \multirow{1}{9em}{$|m_{\ell\ell}-m_Z| < 20$~\GeV}  & $|m_{\ell\ell}-m_Z| > 20$~\GeV  \\
\hline \hline
\end{tabular}
}
\end{center}
\label{table:EventSelection_4l}
\end{table}

The \fivelepton channel requires at least five leptons with at least two SFOS pairs in the event.
The two SFOS lepton pairs with their invariant mass closest to $m_Z$ are identified as \Z boson candidates and must fulfil $|\mll - m_Z| < 20$~\GeV. The remaining lepton is assigned as a lepton from another $Z/W$ boson decay and no additional quality requirements or isolation requirements are applied. Events with at least one $b$-jet are rejected.
Table~\ref{table:EventSelection_5l} summarises the selection criteria.

\begin{table}[h]
\caption{Overview of the criteria used to select inclusive $5\ell$ events, which form the \fivelepton SR.}
\begin{center}
\resizebox{0.64\textwidth}{!}{
\begin{tabular}{c|c}
\hline \hline
\multicolumn{2}{c}{\textbf{Inclusive $5\ell$ event selection ($5\ell$ SR)}} \\
\hline
Satisfy preselection criteria   & {\checkmark}   \\
\multirow{2}{*}{Leptons}  & {At least five leptons}            \\
& {``Loose'' isolation requirement based on tracker and calorimeter measurements}  \\
\Z boson candidates  & {At least two SFOS pairs with {$|m_{\ell\ell}-m_Z| < 20$~\GeV}} \\
$Z$ boson invariant mass      & {$|m_{\ell\ell} - m_Z|<20$~\GeV}               \\
Number of $b$-jets       & {$=0$}                                                 \\
\hline \hline
\end{tabular}
}
\end{center}
\label{table:EventSelection_5l}
\end{table}

To further increase the separation between the signal and background events in the SRs, seven BDT discriminants are trained using the XGBoost package~\cite{Chen:2016btl}\footnote{Julia packages (\texttt{UnROOT.jl}~\cite{UnROOT-2022} and {\texttt{XGBoost.jl}}) were used for the $4\ell$ channel.} and are applied separately to each of the seven SRs. The majority of the input variables are kinematic features of the events.
For the $3\ell$ channel, the BDT is trained with 17 variables for \TLjone, 20 variables for \TLjtwoin, and 24 variables for \TLjtwoout, with some of these variables overlapping between the channels. For the $4\ell$ channel, the same 23 variables are used as inputs for the BDT training for \FLDF, \FLSFZ, and \FLSFnoZ. For the $5\ell$ channel, the BDT is trained with 11 variables. All backgrounds except fake backgrounds (due to non-prompt leptons from hadron decay or jets misidentified as leptons) are included in the BDT training. Since XGBoost cannot handle negative-weight events, the absolute value of each event weight is used.
A five-fold training and cross-validation procedure is used to produce the final discriminant for each SR, and each of the five models is trained on 80\% of the expected signal and background events. Each of the five trained BDTs is applied to the remaining 20\% of the total events. The final BDT score is obtained by averaging the outputs of the five trained BDTs and is used to generate the BDT distribution for the fit.

\subsection{Background estimation }
\label{sec:bkg}
The SM background processes can be divided into two categories: processes with three, four, or at least five prompt leptons in the final state, and fake background processes. %
Depending on the source, the background is estimated in data or simulation, as described below, or a combination of both.

The dominant background source in the $3\ell$ channel is from $WZ+$jets production, followed in importance by $ZZ+$jets production. The $WZ+$jets process contributes 83\% of the expected background in the \TLjone SR, 79\% in the \TLjtwoin SR, and 81\% in the \TLjtwoout SR. The $ZZ+$jets process contributes 11\% of the expected background in the \TLjone SR, 10\% in the \TLjtwoin SR, and 7\% in the \TLjtwoout SR. Due to the presence of three isolated leptons in the final states, the fake background contribution from the $Z+$jets process is $<3$\%. As a result, all backgrounds are estimated with MC simulated events.

The dominant background sources in the $4\ell$ channel are from $ZZ+$jets and fake background processes. The $ZZ+$jets process contributes 50\% of the expected background in the \FLDF SR, 98\% of the events in the \FLSFZ SR, and 93\% in the \FLSFnoZ SR. The fake background mainly originates from the $WZ+$jets and $Z+$jets processes where one or two jets are identified as isolated leptons. The contribution from $t\bar{t}$ production is found to be negligible due to the $b$-jet veto requirement. The fake background contributes 17\% of the expected background in the \FLDF SR, 1\% in the \FLSFZ SR, and 3\% in the \FLSFnoZ SR. The $t\bar{t}Z$ process also contributes 21\% of the expected background in the \FLDF SR.  Due to its contribution to the most sensitive \FLDF SR, the fake background is estimated by using a data-driven method as described in Ref.~\cite{STDM-2014-05}, while all other backgrounds are estimated with MC simulations.

The dominant backgrounds in the $5\ell$ channel originate from fake background processes such as $ZZ+$jets and $t\bar{t}Z$ production. The $ZZ+$jets process contributes 95\% of the expected background in the SR, while the $t\bar{t}Z$ process contributes 2.5\%. These backgrounds have one or two jets misidentified as isolated leptons and are estimated by using a combination of data-driven and simulation-based techniques.

Five control regions (CRs) are defined to check the background modelling in different channels: $WZ+$jets ($3\ell$ channel), $ZZ+$jets ($4\ell$ channel), $Z+$jets ($3\ell$ channel), and two $t\bar{t}Z$ CRs (one for the $3\ell$ channel and the other for the $4\ell$ channel). The two $t\bar{t}Z$ CRs are orthogonal to each other due to the requirement on the number of leptons in the event.

The $WZ+$jets CR is defined with the same set of selection criteria as used for the inclusive $3\ell$ SRs except for a requirement that the BDT score (described in Section~\ref{sec:results}), be less than 0.42. The purity of $WZ$ events in this CR is $\sim 76$\%.
The $ZZ+$jets CR has the same selection criteria as used in the \FLSFZ SR except the \met requirement is reversed to have $\met<10$ \GeV. The purity of $ZZ$ events in this CR is $\sim 99$\%.
In addition, the two $3\ell$ SRs (\TLjone and \TLjtwoin) and the two $4\ell$ SRs (\FLSFZ and \FLSFnoZ) with significant contributions from the $WZ+$jets and $ZZ+$jets processes, help constrain the estimates for these two background processes.

The $Z+$jets CR has the same selection criteria as used in the \TLjone SR except the $W$-lepton is required to have $8<\pT<15$ \GeV, the invariant mass of the three charged leptons is required to be below 150 \GeV, and there is exactly one jet reconstructed in the event and this jet is not tagged as a $b$-jet. The purity of $Z+$jets events in this CR is $\sim 72$\%.

The $t\bar{t}Z$ CR in the $3\ell$ channel is defined in the same way as the signal region, with the exception that at least four jets are required, of which at least two are $b$-tagged. The purity of $t\bar{t}Z$ events in this CR is $\sim 66\%$. The $t\bar{t}Z$ CR in the $4\ell$ channel is defined in the same way as the \FLSFZ SR, with the exception that at least one $b$-jet be present in the event. The purity of $t\bar{t}Z$ in this CR is $\sim 74\%$.

The data-driven method used to estimate the fake background in the $4\ell$ and $5\ell$ channels defines lepton-like jets by requiring the leptons to meet a looser selection criterion but fail to meet the signal-lepton requirement. Compared with the signal leptons, muon-like jets have to satisfy the ``Loose'' quality requirement and electron-like jets have to satisfy the ``Loose'' quality requirement based on the electron likelihood discriminant and have at least one hit in the innermost pixel layer
for the $4\ell$ channel, while both muon-like and electron-like jets are required to satisfy a looser quality requirement for the $5\ell$ channel. In addition, $|d_0|/\sigma_{d_0}$ is required to be less than 10 and no requirements are applied on the isolation variables for both muon-like and electron-like jets.
Events containing up to two lepton-like jets are weighted by a ``fake factor'' to predict the non-prompt lepton background contribution, selected from data for the \fourlepton channel and from MC simulation for the \fivelepton channel. The fake factor is the ratio of the number of non-prompt leptons satisfying the signal lepton criteria over the number satisfying the lepton-like jet criteria. Its value is derived from data samples enriched in $Z+$jets and $t\bar{t}$ production. For the $4\ell$ channel, the dominant fake background originates from the $WZ+$jets process where most jets are light-flavour jets. For the $5\ell$ channel, the dominant fake background originates from both $ZZ+$jets and $t\bar{t}Z$ processes where jets are a mixture of light-flavour and heavy-flavour jets. The fake factor measured in the $Z+$jets sample is thus used for the fake background estimate for the $4\ell$ channel, while the fake factors measured in the $Z+$jets and $t\bar{t}$ samples are combined according to the light-flavour to heavy-flavour background ratio expected in the SR for the $5\ell$ channel.

The normalisations of simulated $WZ+$ jets, $ZZ+$jets, $Z+$jets, and $t\bar{t}Z$ backgrounds are determined from data in the likelihood fits of SRs and CRs described in Section~\ref{sec:results}.

\section{Signal extraction and combination }
\label{sec:results}
The $3\ell$, $4\ell$, and $5\ell$ regions are combined using the profile likelihood method based on a simultaneous fit to distributions in the SRs and CRs. Seven SRs defined in Section~\ref{sec:event_selection} and five CRs defined in Section~\ref{sec:bkg} are used in the fit.
The distributions used in the fit are the seven BDT distributions for the seven SRs and the jet multiplicity distributions in the $WZ+$jets and $ZZ+$jets CRs.
The number of selected events in the $Z+$jets CR, the $t\bar{t}Z$ CR in the $3\ell$ channel, and the $t\bar{t}Z$ CR in the $4\ell$ channel are each included as a single bin in the fit.

For the signal BDT distributions, a total of 58 bins are used: 10 bins for each SR of the $3\ell$ channels, 10 bins collectively for the \FLDF channels, eight bins each for the \FLSFZ and \FLSFnoZ channels, and two bins for the $5\ell$ channel. For the CR distributions, a total of nine bins are used: three bins each for the $WZ$ (1, 2, or $\ge 3$ jets) and $ZZ$ (0, 1, or $\ge 2$ jets) CRs, and one bin each for the $Z+$jets CR, the $t\bar{t}Z$ CR in the $3\ell$ channel, and the $t\bar{t}Z$ CR in the $4\ell$ channel. In total, 67 bins are used in the combined fit.

A binned likelihood function~\cite{Cranmer:2012sba} is
constructed as a product of Poisson probability terms over all bins considered.
This likelihood function depends on the signal-strength parameter $\mu$, a
multiplicative factor that scales the expected number of signal events, and
{\textbf{$\theta$}}, a set of nuisance parameters that encode the effect of
systematic uncertainties in the signal and background expectations~\cite{STDM-2014-05}. The
nuisance parameters are implemented in the likelihood function as Gaussian, log-normal or Poisson constraints, depending on their origin. Correlations of systematic uncertainties arising from common sources are maintained across processes and channels.

The fit includes nine unconstrained parameters that scale the number of events for a particular
process predicted by MC simulation: the signal strength $\mu$, three scale factors ($\mu_{WZ+1\text{~jets}}$, $\mu_{WZ+2\text{~jet}}$, and $\mu_{WZ+\ge 3\text{~jets}}$) for
$WZ+1$ jets, $WZ+2$ jet, and $WZ+\ge 3$ jets, three scale factors ($\mu_{ZZ+0\text{~jets}}$, $\mu_{ZZ+1\text{~jet}}$, and $\mu_{ZZ+\ge 2\text{~jets}}$) for
$ZZ+0$ jets, $ZZ+1$ jet, and $ZZ+\ge 2$ jets, one scale factor ($\mu_{Z+\text{jets}}$) for the $Z+$ jets process,
and one scale factor ($\mu_{t\bar{t}Z}$) for the $t\bar{t}Z$ process.
For the combined fit, the same value for $\mu = \mu_{VVZ}$ is assumed for the \WWZ, \WZZ and \ZZZ processes.
The ratio of on-shell $WVZ$ production to $VH \rightarrow VWW^*/VZZ^*$ production is determined from
MC simulation and is allowed to vary within the theoretical uncertainties of the two processes.

Experimental uncertainties are related to the lepton trigger, reconstruction
and identification efficiencies~\cite{PERF-2017-01, PERF-2015-10}, lepton
isolation criteria~\cite{HIGG-2017-02}, lepton energy (momentum) scale and
resolution~\cite{PERF-2015-10, EGAM-2021-02}, jet energy scale and
resolution~\cite{JETM-2018-05}, jet vertex tagging~\cite{PERF-2014-03},
$b$-tagging~\cite{FTAG-2018-01, FTAG-2020-08, FTAG-2019-02}, modelling of pile-up
and missing transverse momentum~\cite{JETM-2020-03}, and integrated
luminosity~\cite{DAPR-2021-01,LUCID2}.  Nuisance parameters related to these
uncertainties are treated as correlated between all channels.

For each of the background processes evaluated using simulation, a nuisance
parameter representing its normalisation uncertainty is included.
For dominant backgrounds from the $WZ+$jets and $ZZ+$jets processes, the simultaneous fit model has
the power to constrain their normalisations at the $\sim$5\% level.

Uncertainties in data-driven background evaluations mainly come from
statistical and systematic uncertainties in the lepton fake factor measurement.
Additional uncertainties come from the statistical uncertainties of the subsamples used to
extrapolate the background evaluations to the SRs. Nuisance parameters are treated as correlated for backgrounds evaluated using the same method and from the same sources of systematic uncertainty.

Shape-only variations of the signal and simulation-based background distributions due to QCD renormalisation and factorisation scales, PDF, and parton-shower matching scales are considered in the simultaneous fit. The corresponding nuisance parameters for the signal distributions are treated as correlated between the \WWZ, \WZZ and \ZZZ channels.

Table~\ref{tab:postfit_yields} shows the post-fit background, signal, and observed yields for all SRs.
The three most sensitive channels are \TLjtwoout, \FLDF, and $5\ell$.
The fitted scale factors for various background processes are:
$\mu_{WZ+1\text{~jets}}=1.03 \pm 0.11$, $\mu_{WZ+2\text{~jet}}=0.95 \pm 0.16$, $\mu_{WZ+\ge 3\text{~jets}}=0.95 \pm 0.26$,
$\mu_{ZZ+0\text{~jets}}=1.03 \pm 0.10$, $\mu_{ZZ+1\text{~jet}}=0.99 \pm 0.13$, $\mu_{ZZ+\ge 2\text{~jets}}=0.84 \pm 0.24$,
$\mu_{Z+\text{jets}}=0.83 \pm 0.10$, and
$\mu_{t\bar{t}Z}=1.31 \pm 0.17$.
Contributions from SM processes producing the same detector signature as events
in these SRs besides those listed are combined into ``Others''.  The uncertainties shown include both statistical and
systematic uncertainties. Data and predictions agree within uncertainties in all channels.
The contribution of the $VH$ process to the $VVZ$ yield depends on the decay channel and ranges between 3\% (for \FLSFZ) and 50\% (for \TLjone) in the seven SRs.

\begin{table}[t]
\caption{Data and post-fit predicted yields for all SRs. Uncertainties in the predictions include both statistical and systematic uncertainties added in quadrature; correlations among systematic uncertainties are taken into account in the calculation of the total uncertainties.}
\centering
\resizebox{1.\textwidth}{!}{
\begin{tabular}{c D{,}{\pm}{4.4} D{,}{\pm}{4.4} D{,}{\pm}{4.4} D{,}{\pm}{4.4} D{,}{\pm}{4.4} D{,}{\pm}{4.4} D{,}{\pm}{4.4}}
\hline \hline
Signal region
& \multicolumn{1}{c}{\TLjone}
& \multicolumn{1}{c}{\TLjtwoin}
& \multicolumn{1}{c}{\TLjtwoout}
& \multicolumn{1}{c}{\FLDF}
& \multicolumn{1}{c}{\FLSFZ}
& \multicolumn{1}{c}{\FLSFnoZ}
& \multicolumn{1}{c}{$5\ell$} \\
\midrule
$VVZ$ & 104,17 & 99,15& 173,27 & 26.7,4.6 & 18.6,2.1 & 26.8,4.0 & 3.9,0.6 \\
$WZ+$jets & 4271,91 & 932,26 & 2656,81 & \multicolumn{1}{c}{$-$} & \multicolumn{1}{c}{$-$} & \multicolumn{1}{c}{$-$} & \multicolumn{1}{c}{$-$} \\
$ZZ+$jets & 547,46 & 113,14 & 239,27 & 19.7,1.2 & 1447,35 & 383.2,9.9 & \multicolumn{1}{c}{$-$} \\
$Z+$jets & 130,43 & 35,12 & 59,18 & \multicolumn{1}{c}{$-$} & \multicolumn{1}{c}{$-$} & \multicolumn{1}{c}{$-$} & \multicolumn{1}{c}{$-$} \\
$t\bar{t}Z$ & 8.2,1.0 & 35.5,3.1 & 92.5,7.0 & 8.3,0.9 & 1.8,0.2 & 7.0,0.7 & \multicolumn{1}{c}{$-$} \\
Fake & \multicolumn{1}{c}{$-$} & \multicolumn{1}{c}{$-$} & \multicolumn{1}{c}{$-$} & 6.5,2.0 & 14.5,8.5 & 11.6,4.2 & 4.9,0.6 \\
Others & 219,12 & 65.1,5.5 & 221,12 & 4.5,0.4 & 12.2,0.4 & 10.7,0.4 & \multicolumn{1}{c}{$-$} \\
\hline
Total expected & 5280,68 & 1278,28 & 3440,54 & 65.8,5.1 & 1494,34 & 439,10 & 8.9,0.9 \\
Data & \multicolumn{1}{c}{5273} & \multicolumn{1}{c}{1280} & \multicolumn{1}{c}{3423} & \multicolumn{1}{c}{65} & \multicolumn{1}{c}{1513} & \multicolumn{1}{c}{429} & \multicolumn{1}{c}{13} \\
\hline \hline
\end{tabular}
}
\label{tab:postfit_yields}
\end{table}

\begin{figure}[htbp]
\centering
\includegraphics[width=.65\textwidth]{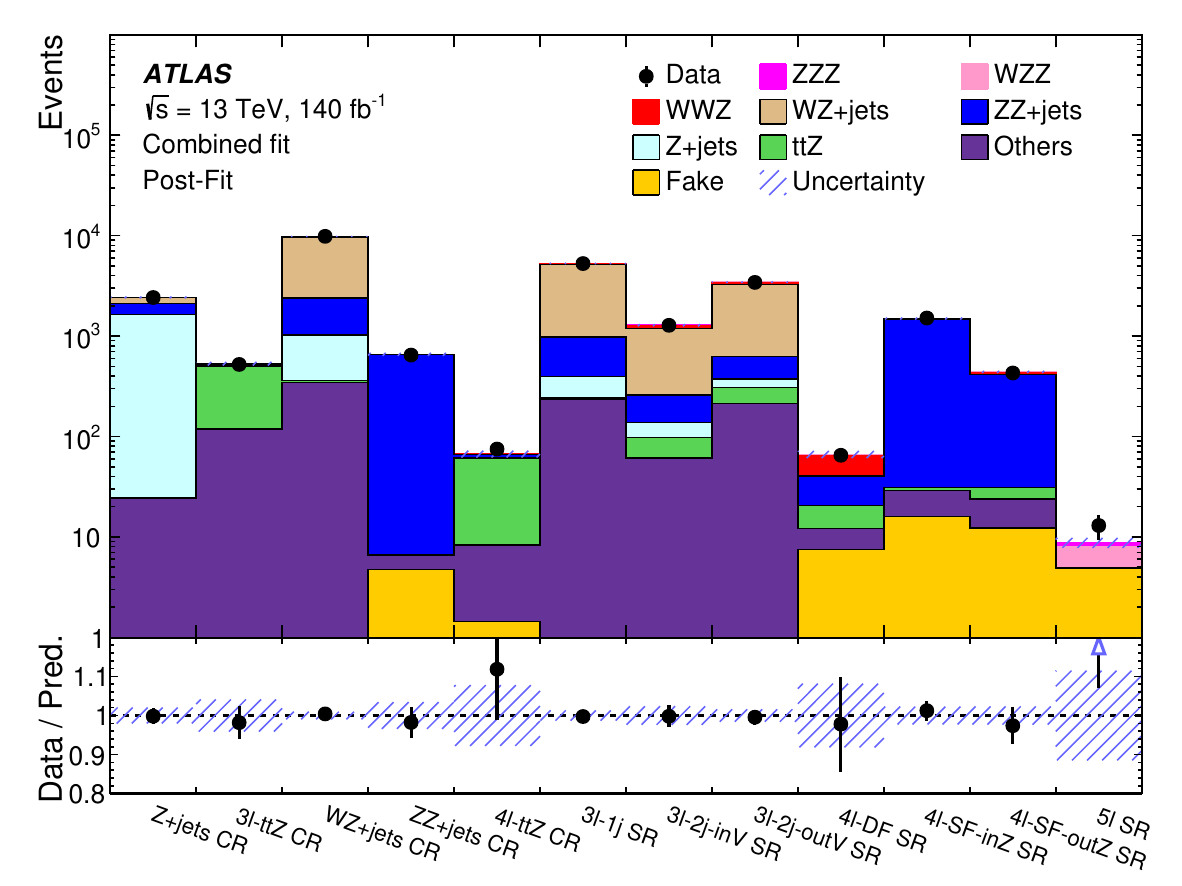}
\caption{Comparison of the observed numbers of events to the predicted yields after fitting for all SRs and CRs. The bottom panel shows the ratio of the data and SM predictions. The uncertainty band includes both statistical and systematic uncertainties obtained by the fit.}
\label{fig:postfit_yields}
\end{figure}

\begin{figure}[htbp]
\centering
\subfloat[]{\includegraphics[width=.33\textwidth]{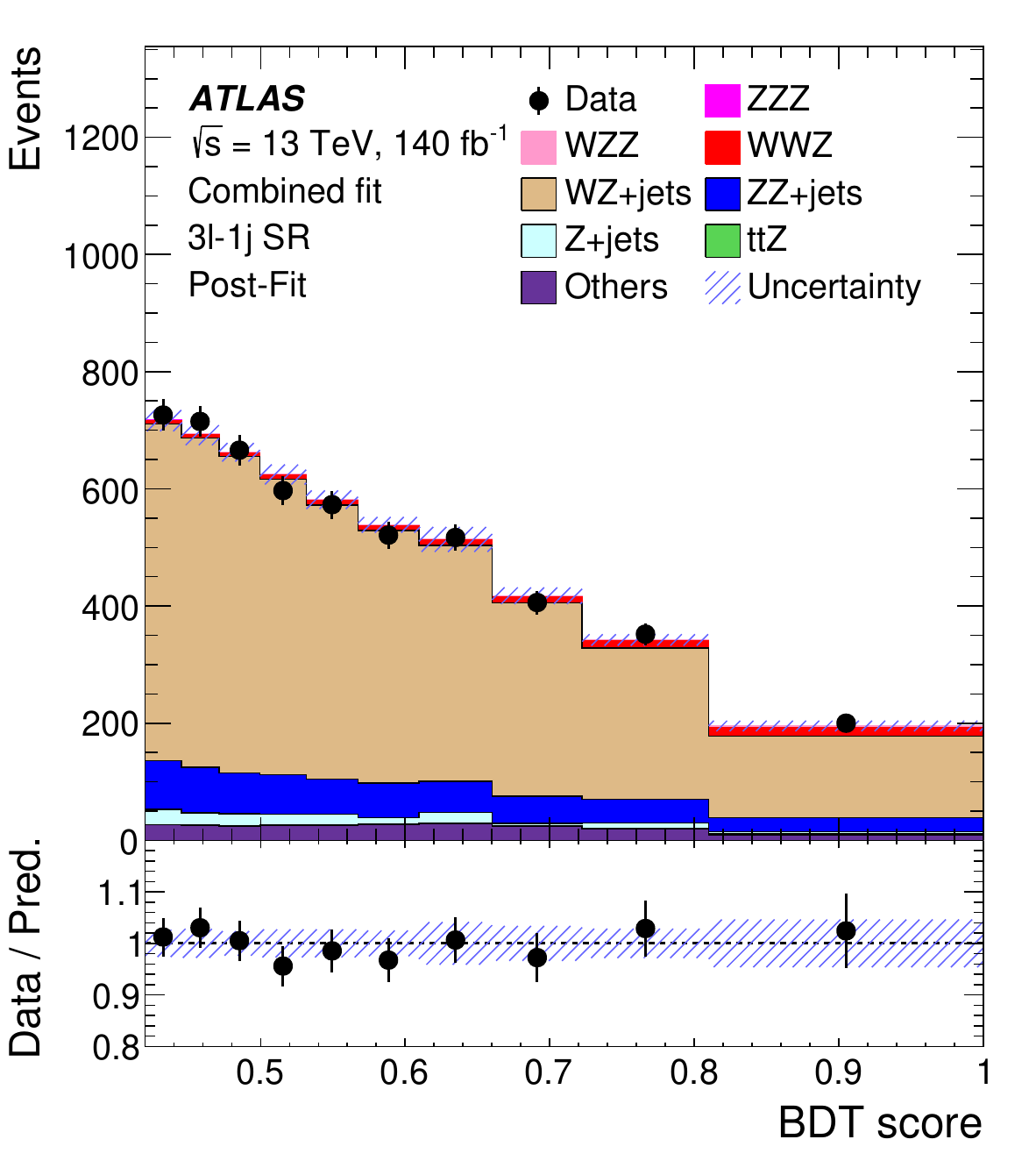}}
\subfloat[]{\includegraphics[width=.33\textwidth]{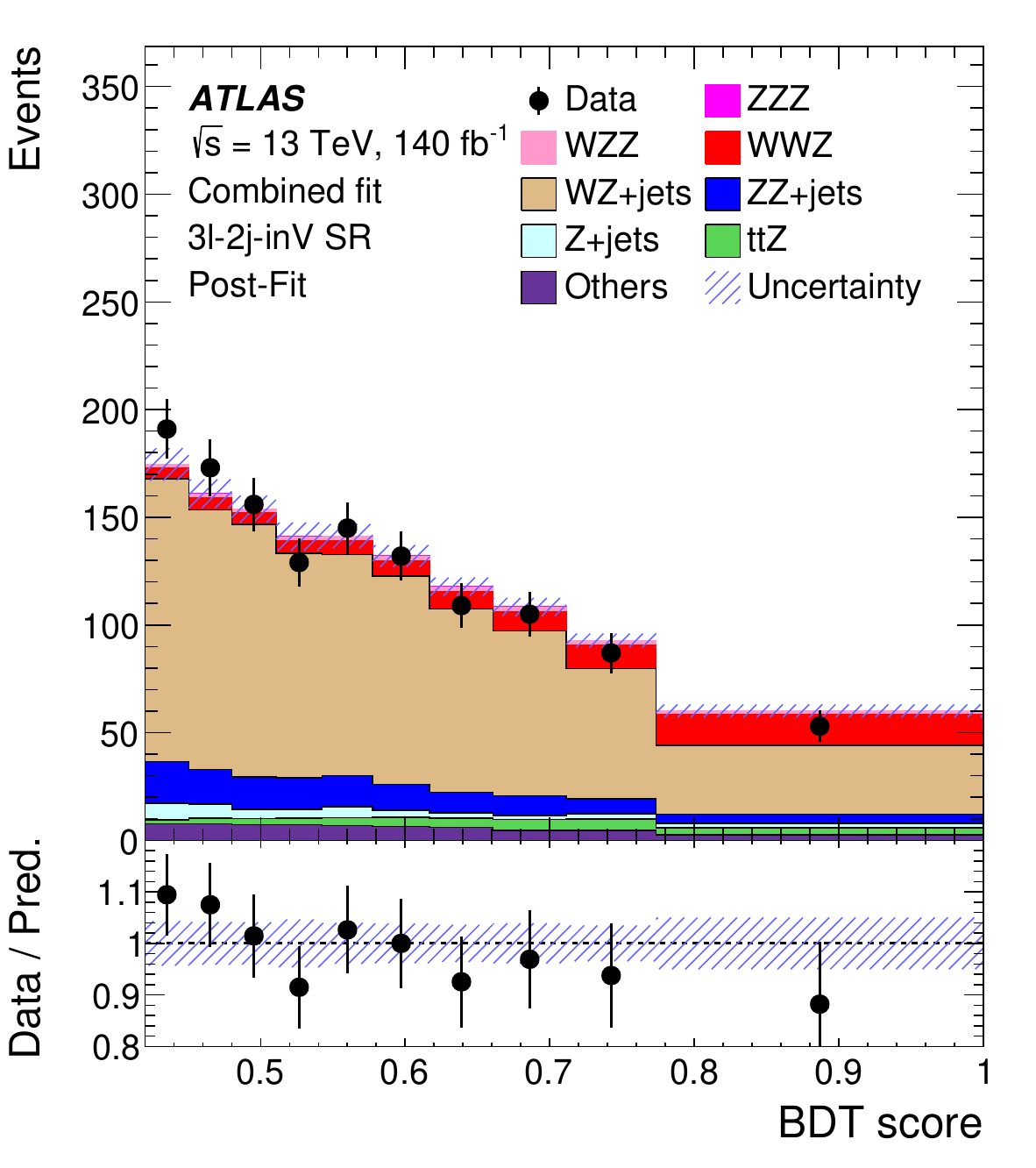}}
\subfloat[]{\includegraphics[width=.33\textwidth]{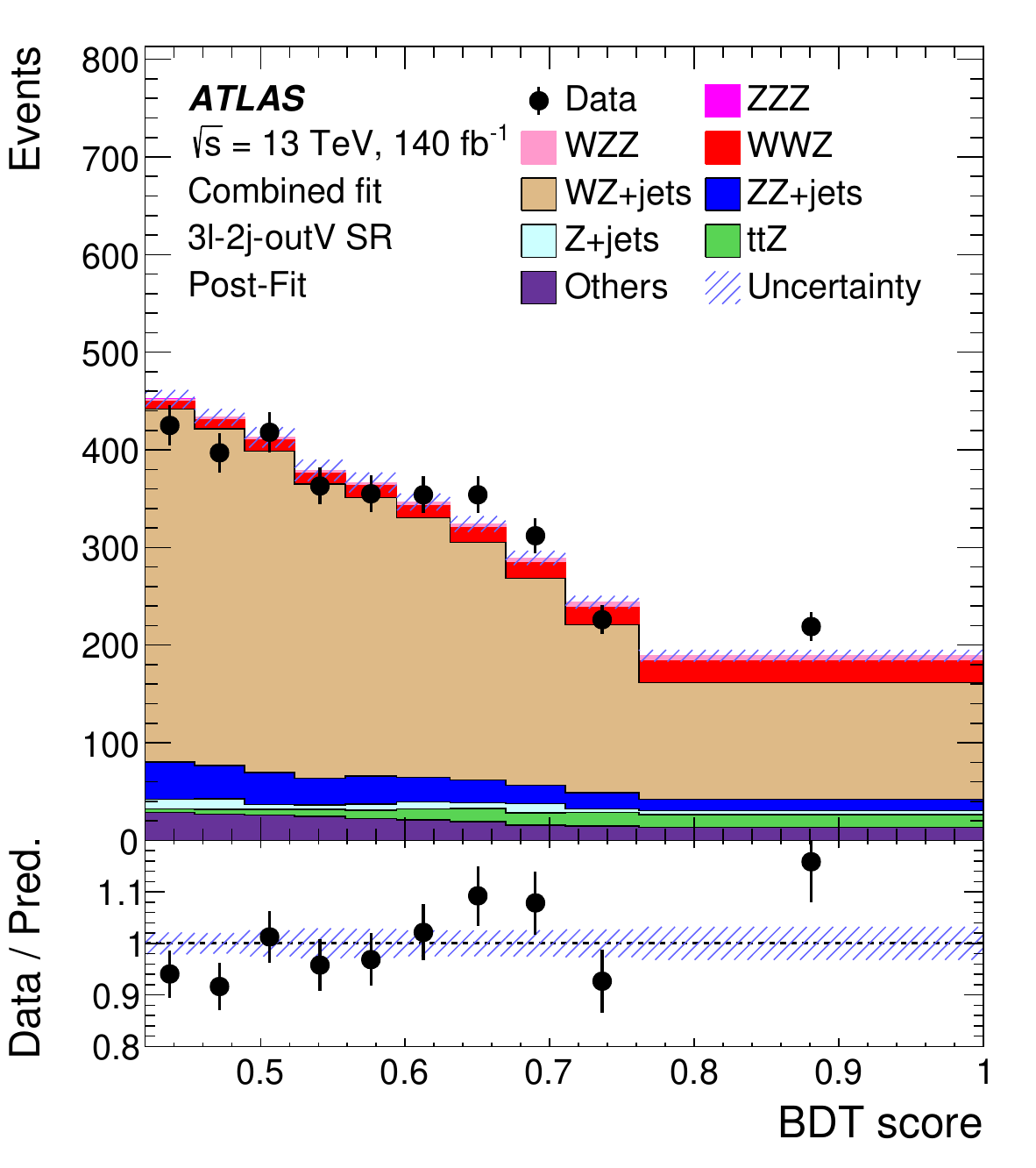}}  \\
\subfloat[]{\includegraphics[width=.33\textwidth]{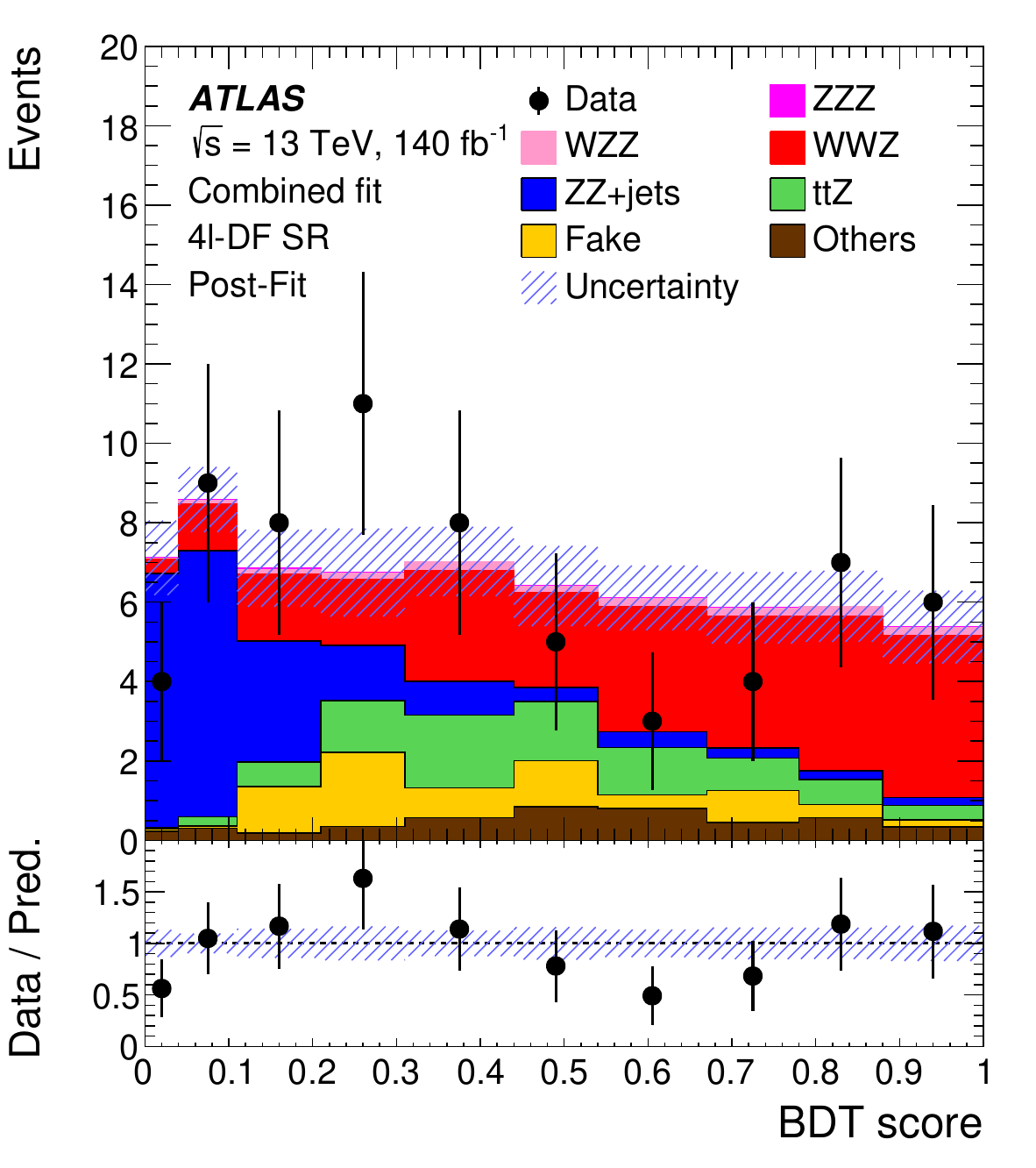}}
\subfloat[]{\includegraphics[width=.33\textwidth]{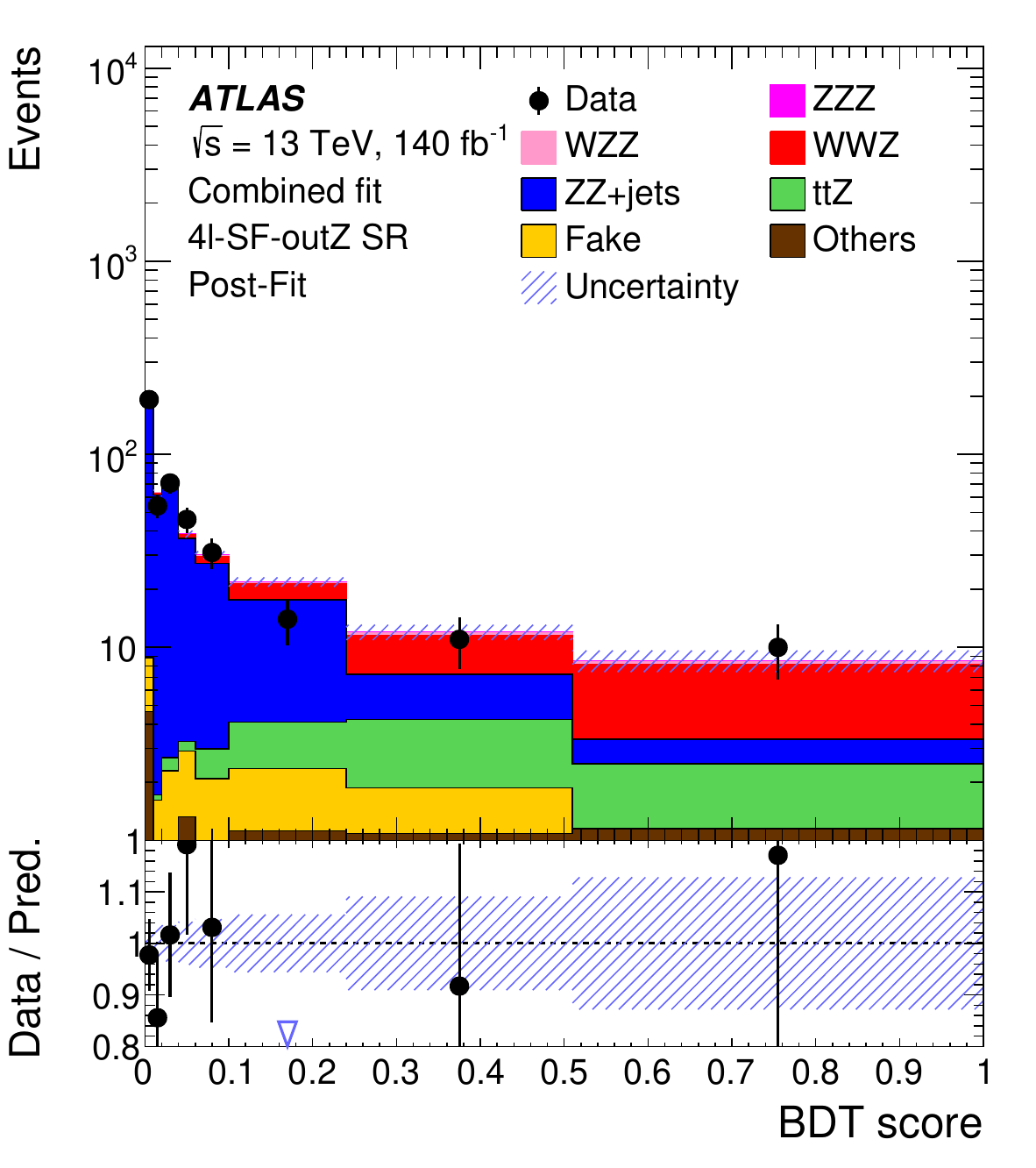}}
\subfloat[]{\includegraphics[width=.33\textwidth]{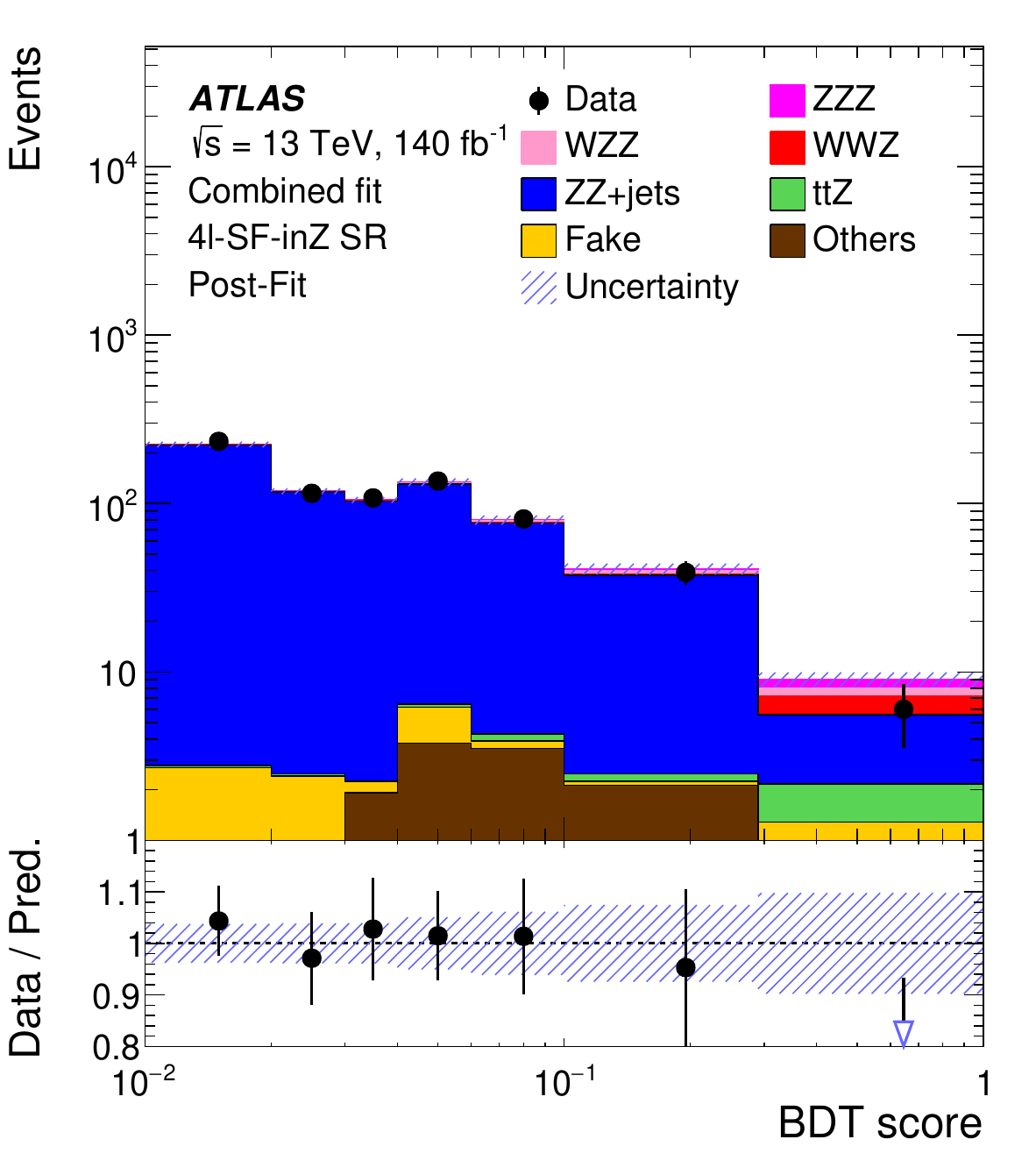}}  \\
\subfloat[]{\includegraphics[width=.33\textwidth]{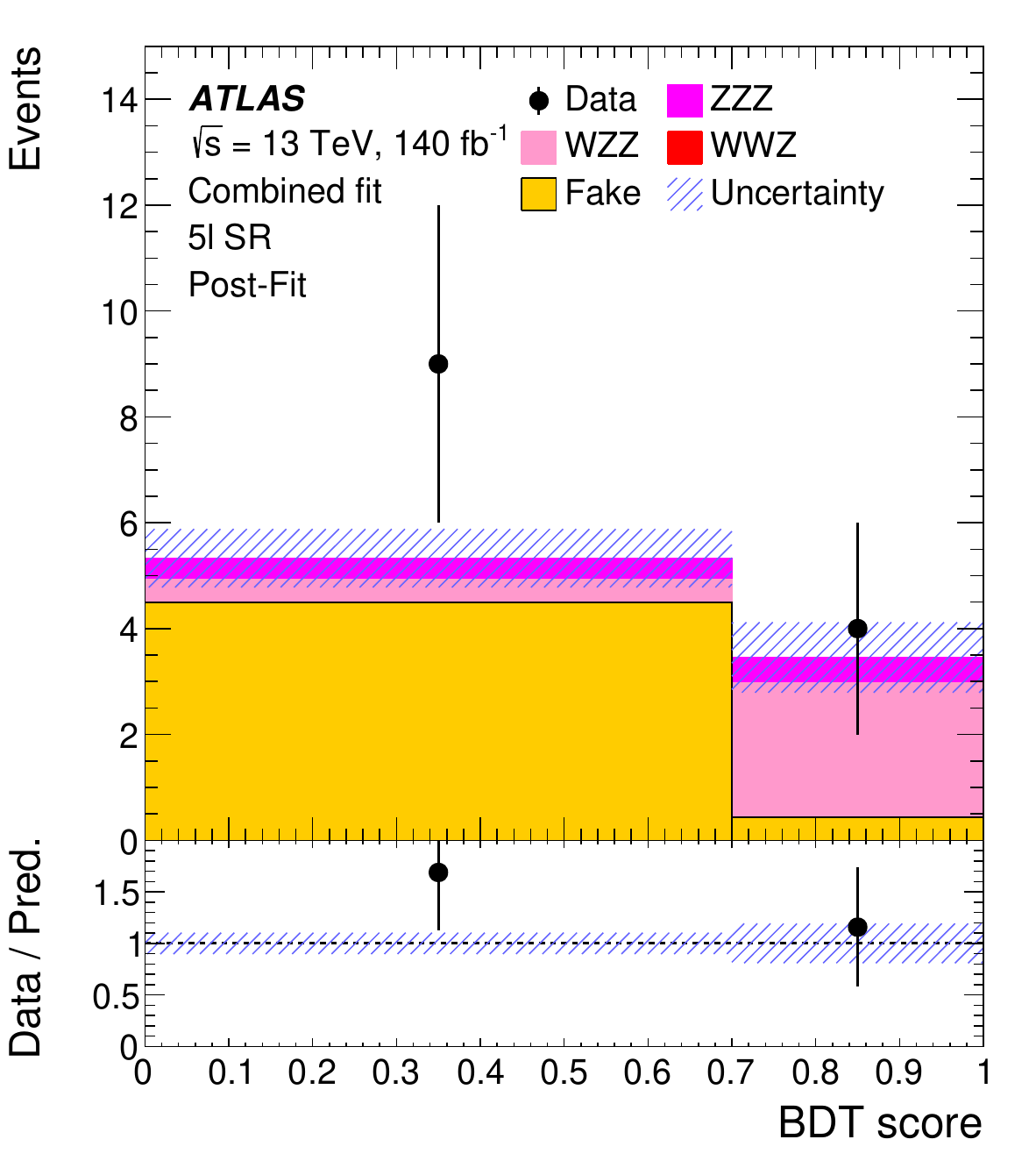}}
\caption{Comparison of the BDT distribution between data and predictions for the (a) \TLjone, (b) \TLjtwoin, (c) \TLjtwoout, (d) \FLDF, (e) \FLSFnoZ, (f) \FLSFZ, and (g) $5\ell$ SRs. The bottom panel shows the ratio of the data and SM predictions. The uncertainty band includes both statistical and systematic uncertainties obtained by the fit.}
\label{fig:postfit_BDT}
\end{figure}

Figure~\ref{fig:postfit_yields} shows the comparison of the observed number of events to the predicted yields after fitting
for both SRs and CRs. Figure~\ref{fig:postfit_BDT} shows the comparison of the BDT distribution between data and predictions for all seven SRs.
The two variables with the highest discriminating power are: $H_T^{\text{tot}}$ (scalar sum of the $\pT$ of leptons and jets) and the second leading jet $\pT$ for the \TLjtwoout channel, the invariant mass of the second $Z$-lepton pair and the jet multiplicity for the \FLDF channel, and the \met significance and the $W$-lepton $\pT$ for the $5\ell$ channel.

Additional fits are performed separately to determine the signal strength of the $WWZ$ process ($\mu_{WWZ}$) and the $WZZ$ process ($\mu_{WZZ}$).
Due to limited statistics, there are no specific SRs defined for $ZZZ$ production, and thus no separate fit is performed to determine the signal strength of the $ZZZ$ process. For these fits the other signal strength is fixed to its SM expectation.

The measured signal strengths combined with the SM predicted cross sections are used to derive the measured cross sections of various processes.
The combined observed (expected) signal strength for the $VVZ$ process is $\mu_{VVZ} = 1.43 \pm 0.20~(\text{stat.})^{+0.21}_{-0.19} (\text{syst.})$ $(1.00^{+0.27}_{-0.25})$. The measured cross section is found to be $660^{+93}_{-90}(\text{stat.})^{+88}_{-81}(\text{syst.})$ fb and the observed (expected) significance corresponds to 6.4 (4.7) $\sigma$, marking the observation of $VVZ$ production.
The observed (expected) significance of $WWZ$ production is 4.4 (3.6) $\sigma$, representing the evidence of this process at the LHC.
Table~\ref{tab:xsections_sensitivity} shows the measured signal strengths, cross sections and observed (expected) sensitivities for the $VVZ$, $WWZ$, and $WZZ$ processes. The systematic uncertainties in the measured signal strengths are dominated by QCD scale uncertainties in the signal processes, uncertainties in reconstructed jet energy scale, resolution and $b$-tagging, and limited MC statistics of the signal samples.

If $VH$ production is considered as part of the background, the combined observed cross section is found to be $\sigma(pp \rightarrow VVZ)=382^{+65}_{-63}(\text{stat.})^{+57}_{-60}(\text{syst.})$ fb with an observed signal strength of $1.59^{+0.24}_{-0.29}~(\text{stat.})^{+0.30}_{-0.25}~(\text{syst.})$. The observed (expected) significance corresponds to 5.5 (3.7) $\sigma$. The ratio of on-shell $VVZ$ production to $VH \rightarrow VWW^*/VZZ^*$ production is determined from MC simulation and is allowed to vary within the theoretical uncertainties of the two processes.

\begin{table}[t]
\captionof{table}{Measured signal strengths and inclusive cross sections and observed (expected) sensitivities for $WWZ$, $WZZ$, and $VVZ$ production. The uncertainties listed are statistical and systematic.}
\centering
\resizebox{1.\textwidth}{!}{
\begin{tabular}{c|ccc}
\toprule \toprule
Process & Signal strength & Cross section (fb) &  Observed (expected) sensitivity \\\hline
\rule{0pt}{14pt} $VVZ$ & $1.43 \pm 0.20({\text{stat.}})^{+0.21}_{-0.19}({\text{syst.}})$ & $660^{+93}_{-90}(\text{stat.})^{+88}_{-81}(\text{syst.})$ & 6.4 (4.7) $\sigma$ \\
\rule{0pt}{14pt} $WWZ$ & $1.33 \pm 0.28({\text{stat.}})^{+0.21}_{-0.17}({\text{syst.}})$ & $442 \pm 94(\text{stat.})^{+60}_{-52}(\text{syst.})$ & 4.4 (3.6) $\sigma$ \\
\rule{0pt}{14pt} $WZZ$ & $2.13^{+1.18}_{-0.96}({\text{stat.}})^{+0.76}_{-0.41}({\text{syst.}})$  & $200^{+111}_{-91}(\text{stat.})^{+65}_{-37}(\text{syst.})$ & 2.8 (1.6) $\sigma$ \\
[0.1cm]
\bottomrule \bottomrule
\end{tabular}
}
\label{tab:xsections_sensitivity}
\end{table}

\section{EFT analysis }
\label{s:eft}
The production of three massive vector bosons allows to study the quartic vector boson vertex. The EFT approach was chosen to constrain physics beyond the SM at higher mass scales. The EFT extends the SM Lagrangian with additional terms of higher dimensions,
\begin{equation}
\mathcal{L}_{S M E F T}=\mathcal{L}_{S M}+\mathcal{L}^{(5)}+\mathcal{L}^{(6)}+\mathcal{L}^{(7)}+\ldots,
\quad \text{ where } \mathcal{L}^{(d)}=\sum_{i=1}^{n_{d}} \frac{f_{i}^{(d)}}{\Lambda^{d-4}} \mathcal{O}_{i}^{(d)} \quad \text{ for } d>4,
\end{equation}
with $f_{i}$ the Wilson coefficient, $d$ the dimension of the operator, $\Lambda$ an arbitrary energy scale set to 1 \TeV, and $\mathcal{O}_{i}$ the EFT operator extending the SM. The Eboli parameterisation~\cite{PhysRevD.74.073005} is used, where the lowest dimension operators acting on the quartic gauge boson vertex are of dimension 8 with the assumption of a light Higgs boson and thus a linear realisation of the $SU(2)_L \otimes U(1)_Y$ symmetry breaking~\cite{PhysRevD.74.073005}.
The Wilson coefficients of the most sensitive operators to massive tri-boson final states are $f_{M2}, \, f_{M3}, \, f_{M4}, \, \text{and} \, f_{M5}$.  One dimensional limits are derived by allowing only one coefficient to vary from zero and setting constraints using
multivariate techniques, combining the \threelepton and \fourlepton channels in a simultaneous fit.
To account for QCD NLO corrections a $k$-factor is applied to the EFT MC samples.
To increase the sensitivity to EFT effects, a BDT is trained on events with contributions from all EFT operators under consideration. The training is repeated for each signal region in the \threelepton and \fourlepton channels, following the same approach as described in Section~\ref{sec:event_selection}.

\begin{figure}[h]
\centering
\subfloat[]{
\includegraphics[width=0.45\textwidth,height=0.25\textheight,keepaspectratio]{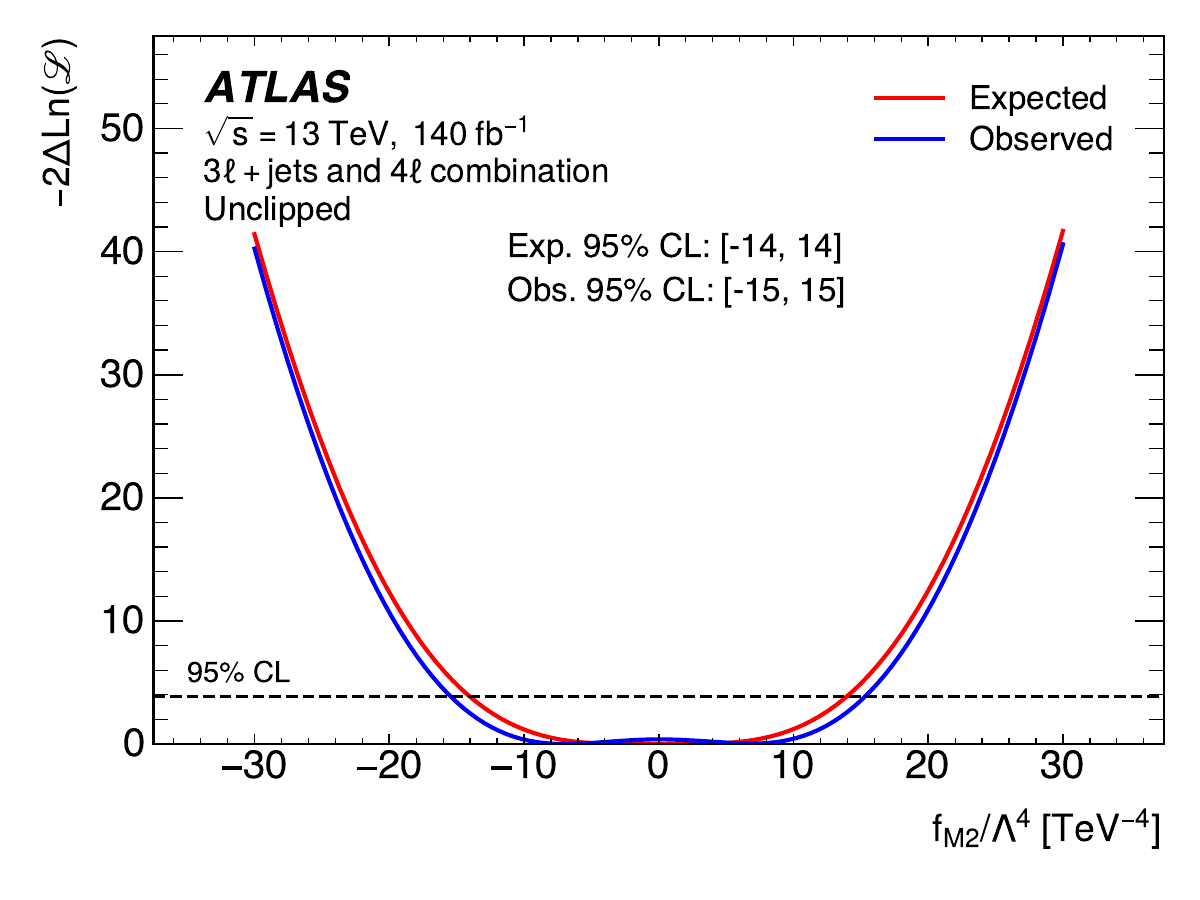}
}
\subfloat[]{
\includegraphics[width=0.45\textwidth,height=0.25\textheight,keepaspectratio]{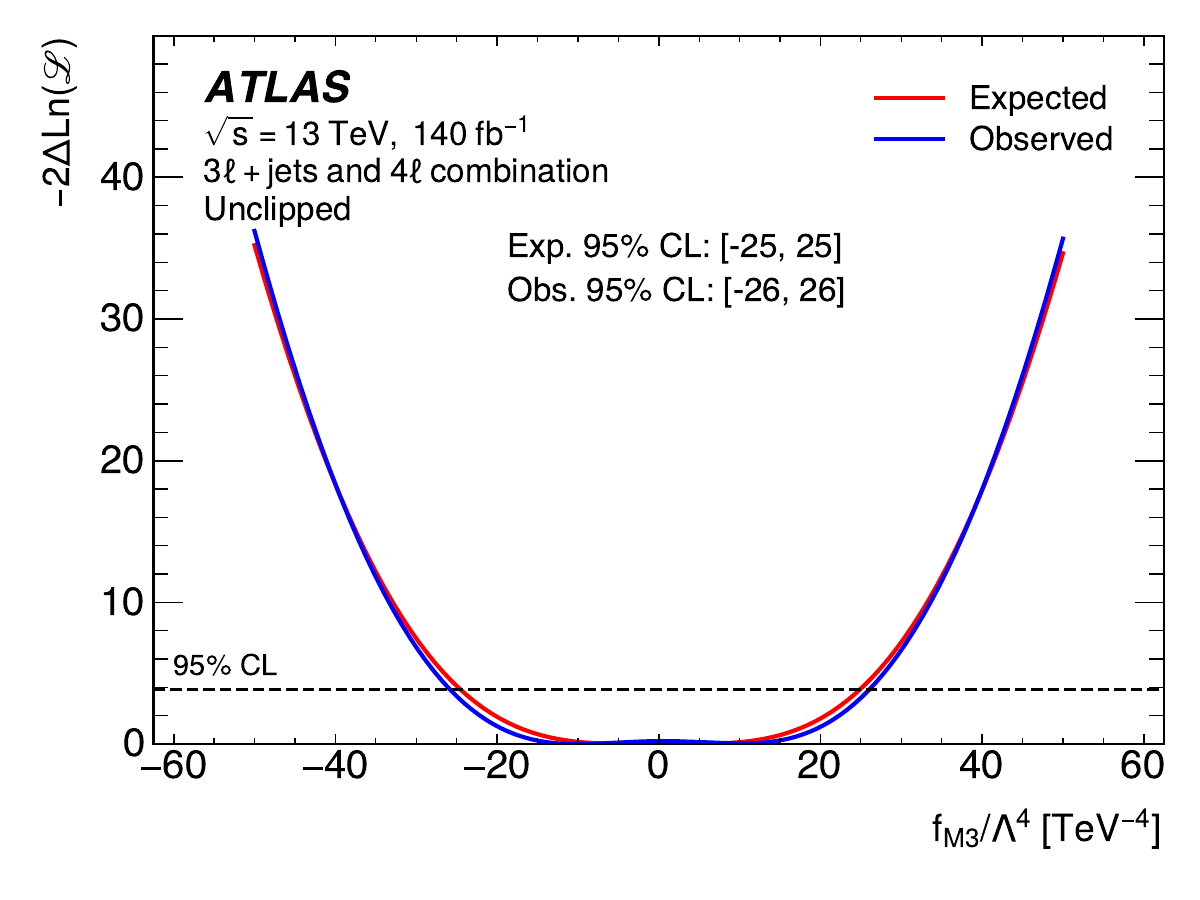}
}
\\
\subfloat[]{
\includegraphics[width=0.45\textwidth,height=0.25\textheight,keepaspectratio]{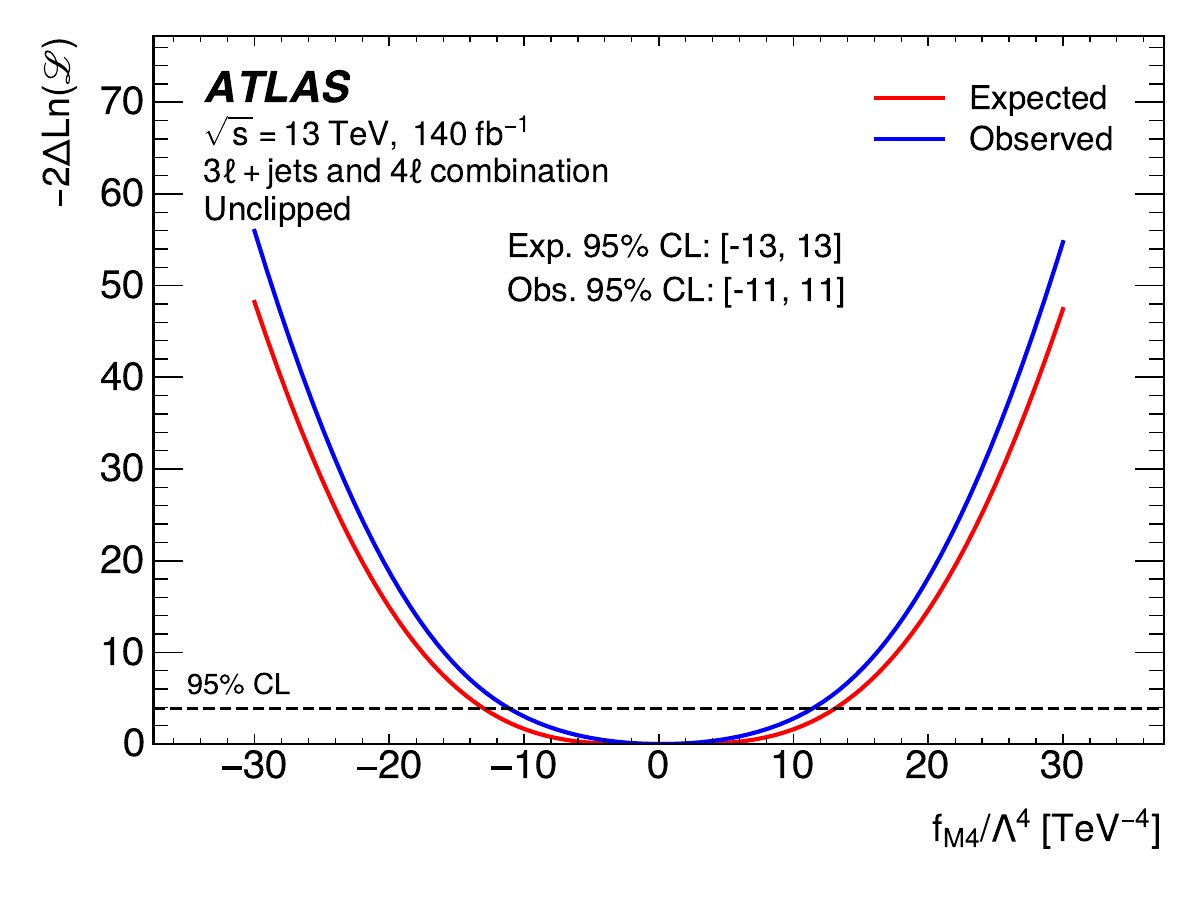}
}
\subfloat[]{
\includegraphics[width=0.45\textwidth,height=0.25\textheight,keepaspectratio]{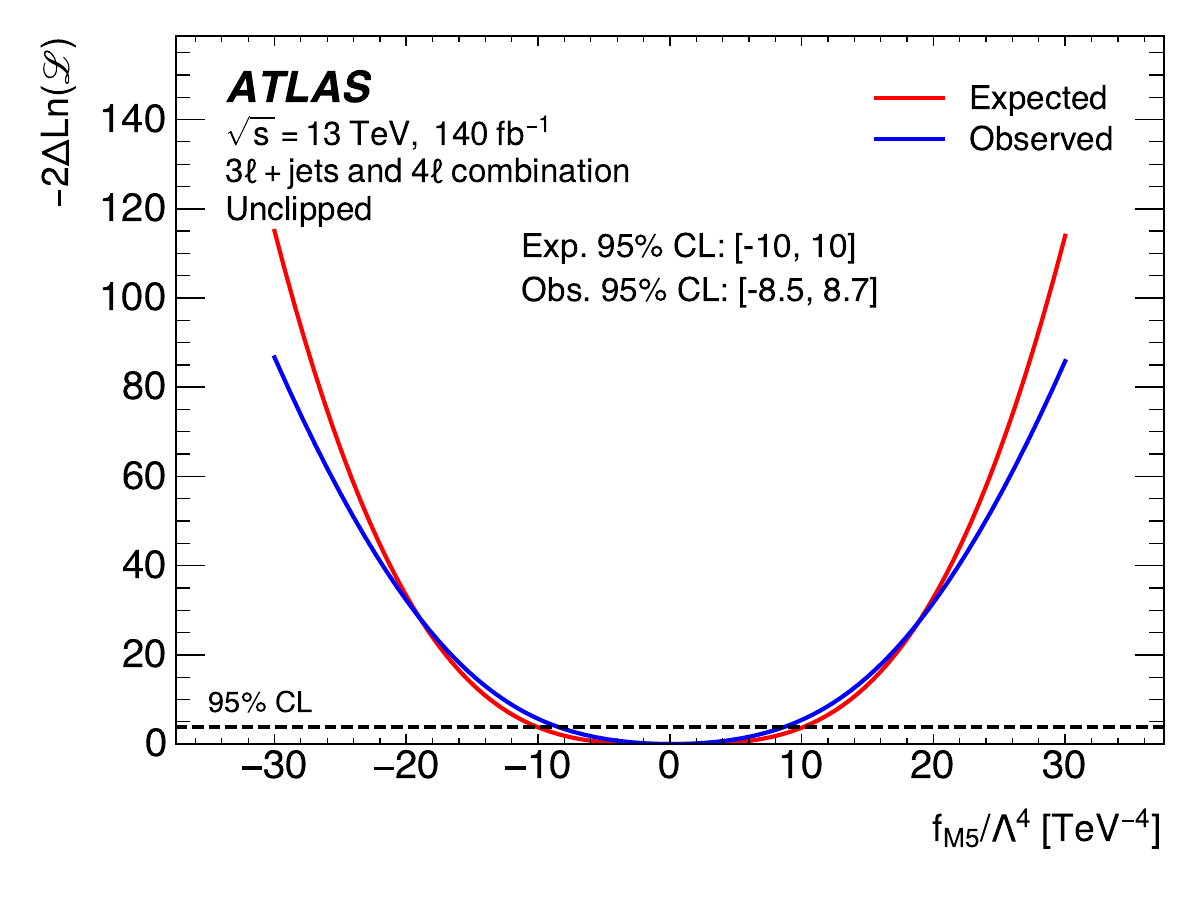}
}
\caption{ The combined 3$\ell$+jets and 4$\ell$ channel log-likelihood curves for the Wilson coefficients (a) $f_{M2}$, (b) $f_{M3}$, (c) $f_{M4}$, and (d) $f_{M5}$. Expected and observed log-likelihood curves are shown, no unitarisation is applied.}
\label{fig:Loglikelihoods}
\end{figure}

The EFT expectation is parameterised as
\begin{equation}
N_{b,H}(\textbf{f},\textbf{$\theta$})=\sum_p \sum_d N_b^{p d,\mathrm{SM}}(\textbf{$\theta$})\left(1+\sum_i A_{b i}^{p d} f_i+\sum_i B_{b i}^{p d} f_i^2+\sum_{i<j} C_{b i j}^{p d} f_i f_j\right)+N_b^{\mathrm{bkg},\mathrm{SM}}(\textbf{$\theta$})\, ,
\end{equation}
with expected number of events, $N_{b,H}$ for a bin $b$ with Wilson coefficients $\textbf{f}$, and the nuisance parameters \textbf{$\theta$}.
The expected number of SM events in tri-boson process $p$ and decay channel $d$ is $N_b^{p d,\mathrm{SM}}(\textbf{$\theta$})$, with expected number of background events $N_b^{\mathrm{bkg},\mathrm{SM}}(\textbf{$\theta$})$. The EFT contributions are modelled with $A_{b i}^{p d}, \, B_{b i}^{p d} \text{ and } C_{b i j}^{p d}$~\cite{ATL-PHYS-PUB-2022-037}.
The likelihood is constructed as
\begin{equation}
L(\boldsymbol{N} \mid \boldsymbol{f}, \boldsymbol{\theta})=\prod_b^{n_{\text{bins }}}
\left(
\frac{N_{b,H}^{N_b} e^{N_{b,H}}}{ N_b!}
\right)  \times \prod_{i=1}^{n_{\text{sys }}} \xi_i\left(\theta_i\right)\, ,
\end{equation}
with \textbf{$N$} signal events per bin over a total number of bins $n_{\text{bins }}$ and $\xi_i$ the constraints on each nuisance parameter, $n_{\text{sys }}$, to account for the experimental systematic uncertainties.
The one dimensional 95\% confidence level (CL) limits on the Wilson coefficients $f_i$ are then extracted using a profile likelihood ratio test allowing only the coefficient under consideration to vary while setting the other coefficients to zero.
The expected and observed 95\% CL limits for the \threelepton and \fourlepton channel separately and combined, including statistical and systematic uncertainties, are shown in Table~\ref{tab:AQGCfullsystlims}. The combined log-likelihood curves for the Wilson coefficients $f_{M2}$, $f_{M3}$, $f_{M4}$, and $f_{M5}$ are shown in Figure~\ref{fig:Loglikelihoods}.

\begin{figure}[h]
\centering
\subfloat[]{\includegraphics[width=.45\textwidth]{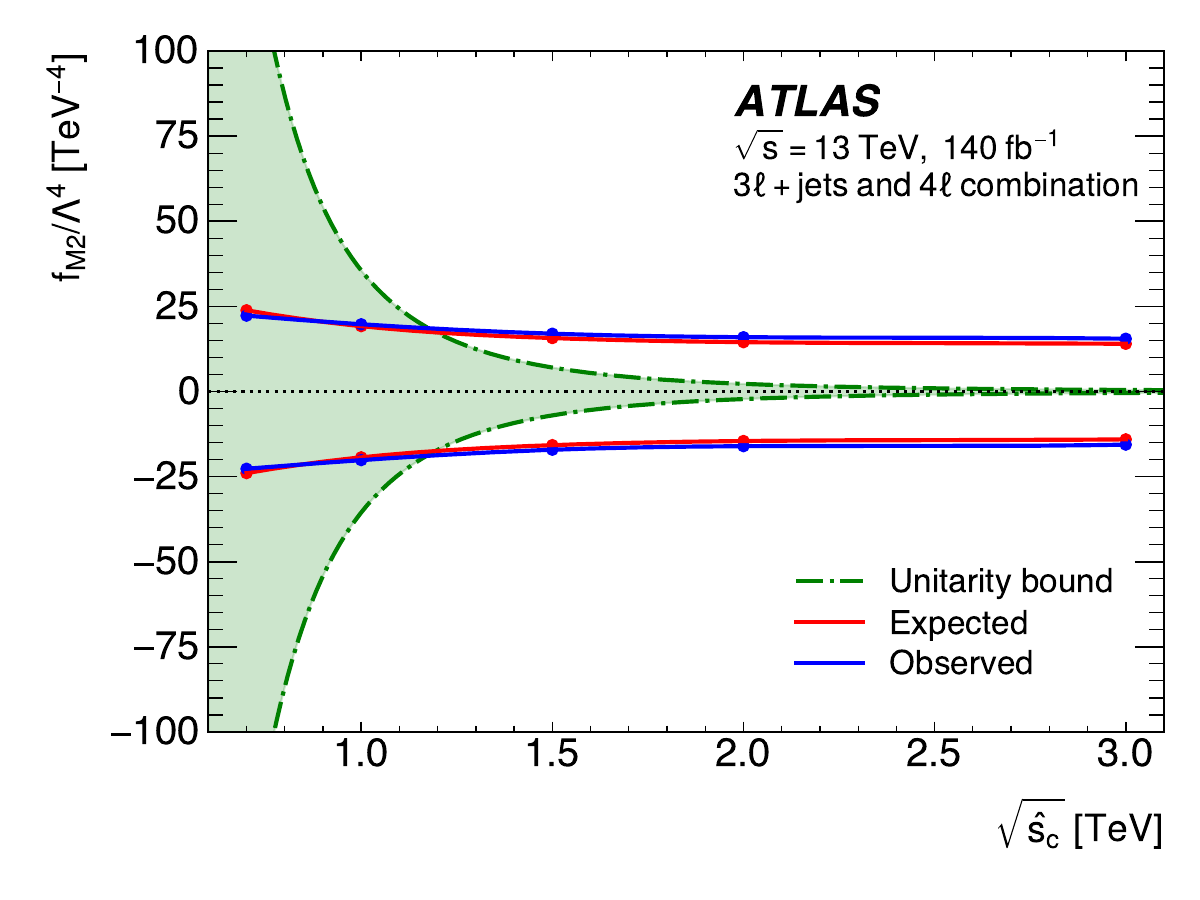}}
\subfloat[]{\includegraphics[width=.45\textwidth]{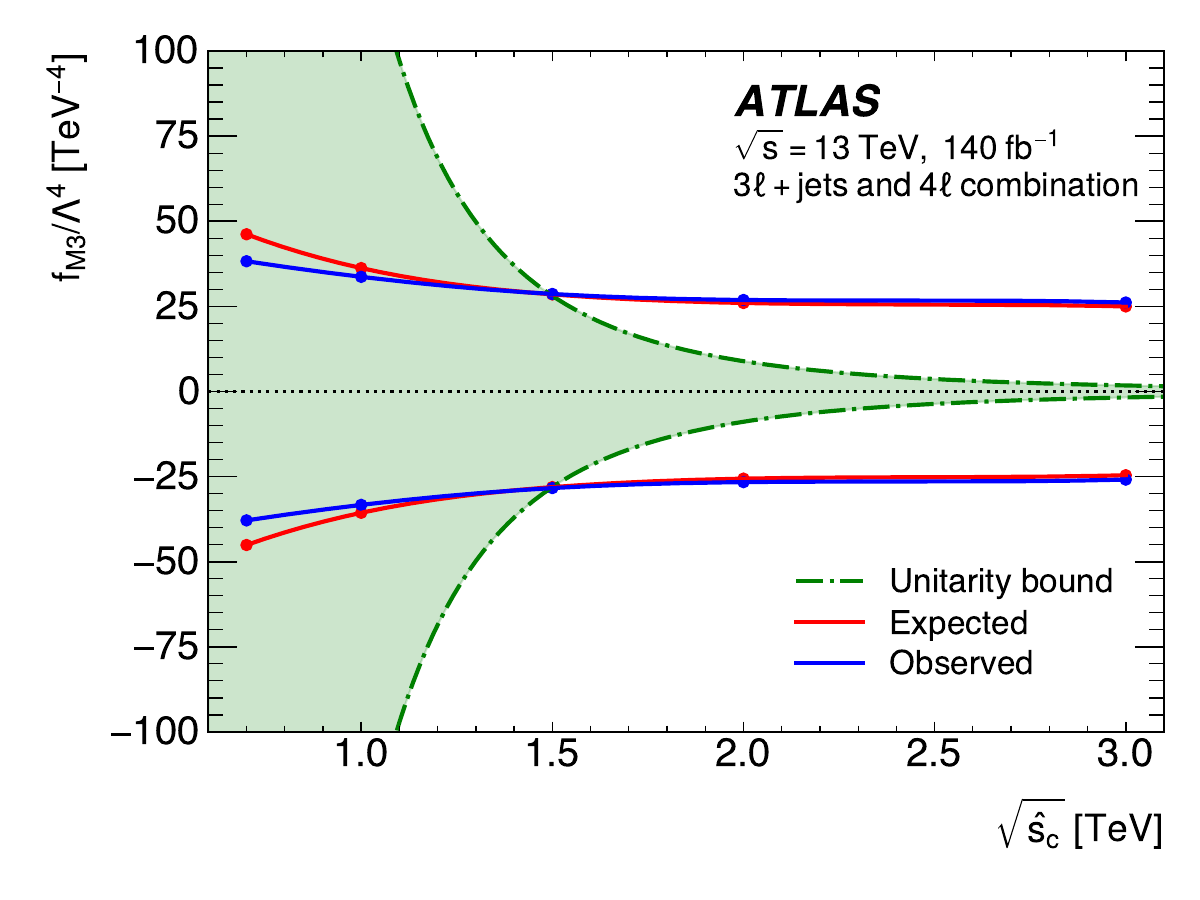}} \\
\subfloat[]{\includegraphics[width=.45\textwidth]{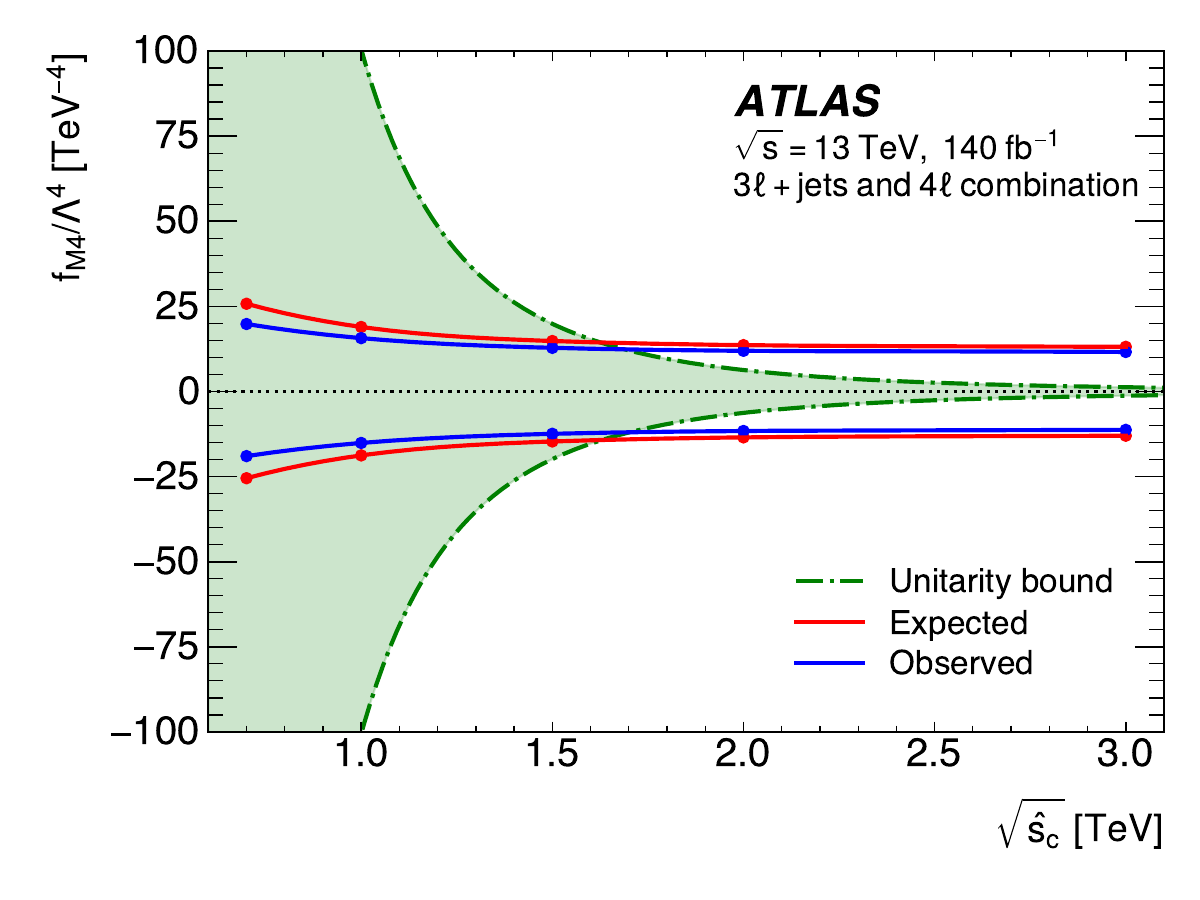}}
\subfloat[]{\includegraphics[width=.45\textwidth]{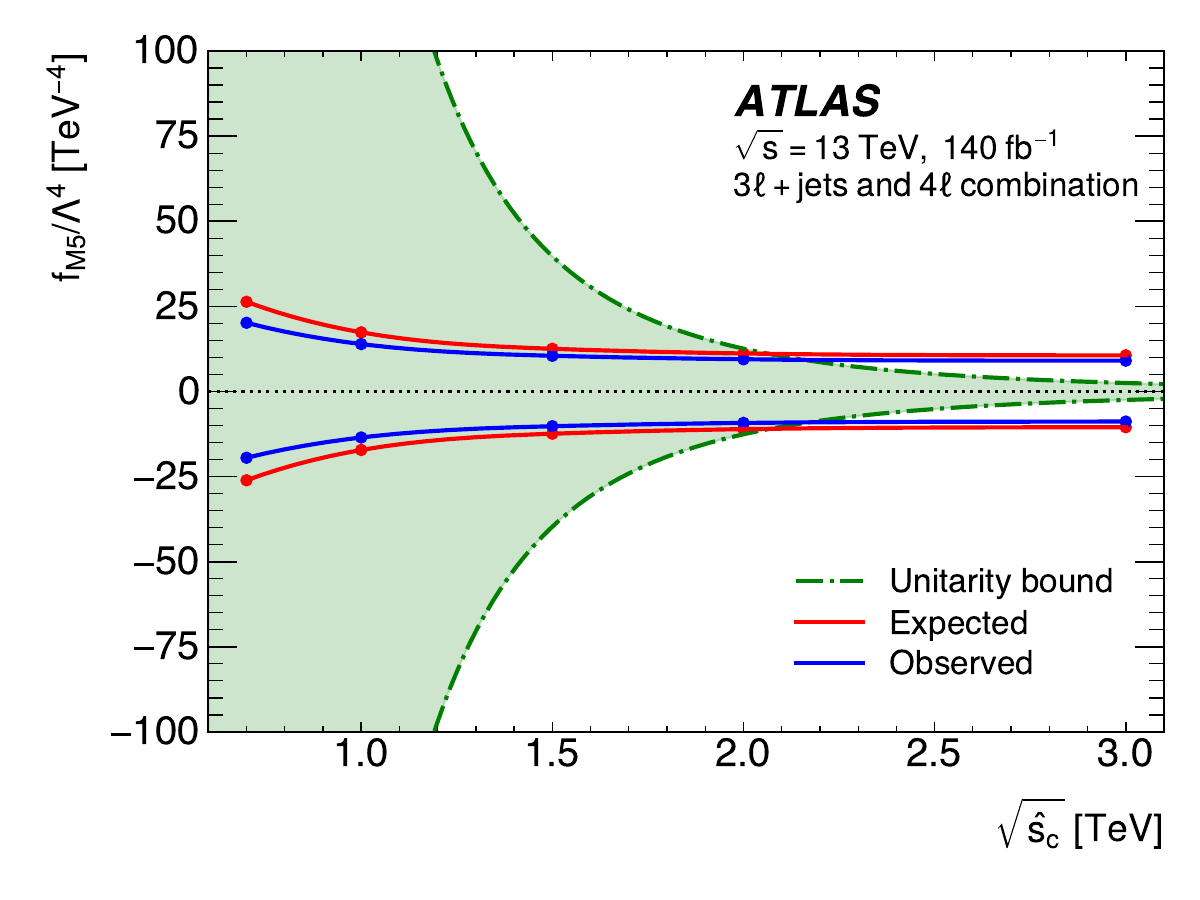}}
\caption{\label{fig:EFTclips} The clipping scans for the Wilson coefficients (a) $f_{M2}$, (b) $f_{M3}$, (c) $f_{M4}$, and (d) $f_{M5}$ are shown together with the unitarity limits as a function of the clipping parameter $\sqrt{\hat{s}_c}$, where the EFT contribution in events with $m_{max}(V_i V_j) > \sqrt{\hat{s}_c}$ is set to zero.}
\end{figure}

The unitarity bounds of the Wilson coefficients were estimated by adapting the existing calculation for dimension-8 operators in two-to-two scattering process into vector boson scattering processes~\cite{Almeida:2020ylr} and setting the parton centre-of-mass energy $\sqrt{\hat{s}}$ to the maximum of the three diboson invariant mass combinations $m_{max}(V_i V_j)$. The limits are then derived as a function of the clipping parameter $\sqrt{\hat{s}_c}$, where the EFT contribution in events with $m_{max}(V_i V_j) > \sqrt{\hat{s}_c}$ is set to zero. The clipping scans~\cite{Szleper:2020WV} for the Wilson coefficients $f_{M2}$, $f_{M3}$, $f_{M4}$, and $f_{M5}$  are shown in Figure~\ref{fig:EFTclips} together with the unitarity limits. The most constraining limit respecting unitarity is found at the intersection of the calculated limit with the unitarity bound, and listed for all operators in Table~\ref{tab:clipcomblim}. These constraints are comparable to published limits derived in the $W\gamma jj$ final state~\cite{ATLAS:2024bho, CMS-SMP-21-011}.

\begin{table}[t]
\caption{The observed (expected) non-unitarised limits on Wilson coefficients for the 3$\ell$ and 4$\ell$ channels separately, and the combined limits.}
\centering
\resizebox{0.75\textwidth}{!}{
\begin{tabular}{l|c|c|c}
\hline
\hline
\multicolumn{4}{c}{Observed (expected) 95\% CL limits on Wilson coefficients ($\textrm{\TeV}^{-4}$)} \\ [0.6ex]
\hline
Coefficient & 3$\ell$ & 4$\ell$ & Combination \\
\hline
$f_{M2}/\Lambda^4$  & $[-15, 15]~([-17, 17])$ & $[-23, 23]~([-18, 18])$ & $[-15, 15]~([-14, 14])$ \\
$f_{M3}/\Lambda^4$  & $[-25, 25]~([-29, 30])$ & $[-39, 40]~([-31, 31])$ & $[-26, 26]~([-25, 25])$ \\
$f_{M4}/\Lambda^4$  & $[-13, 14]~([-16, 16])$ & $[-17, 17]~([-16, 16])$ & $[-11, 11]~([-13, 13])$ \\
$f_{M5}/\Lambda^4$  & $[-11, 11]~([-13, 13])$ & $[-12, 12]~([-13, 13])$ & $[-8.5, 8.7]~([-10, 10])$ \\
\hline
\hline
\end{tabular}
}
\label{tab:AQGCfullsystlims}
\end{table}

\begin{table}[t]
\caption{The observed and expected unitarised limits on Wilson coefficients for the combination of the 3$\ell$ and 4$\ell$ channels on dimension 8 operator Wilson coefficients, together with the energy scale at which the unitarity bound is crossed.}
\centering
\resizebox{0.9\textwidth}{!}{
\begin{tabular}{c|c|c|c|c}
\hline
\hline
Coefficient & Expected limit [$\textrm{\TeV}^{-4}$] & Exp. $\sqrt{\hat{s}_c}$ [\TeV] & Observed limit [$\textrm{\TeV}^{-4}$] & Obs. $\sqrt{\hat{s}_c}$ [\TeV] \\ [0.6ex]

\hline
$f_{M2}/\Lambda^4$  & $[-18, 17]$ & $1.2$ & $[-19, 19]$ & $1.2$  \\
$f_{M3}/\Lambda^4$  & $[-28, 29]$ & $1.5$ & $[-28, 29]$ & $1.5$  \\
$f_{M4}/\Lambda^4$  & $[-14, 14]$ & $1.6$ & $[-12, 12]$ & $1.7$  \\
$f_{M5}/\Lambda^4$  & $[-11, 11]$ & $2.1$ & $[-9.1, 9.3]$ & $2.2$ \\
\hline
\hline
\end{tabular}
}
\label{tab:clipcomblim}
\end{table}

\FloatBarrier

\section{Conclusions}
A search for the joint production of three massive vector bosons ($WWZ$, $WZZ$, and $ZZZ$) is presented using 140 fb$^{-1}$ of data at $\sqrt{s} = \SI{13}{\TeV}$ collected by the ATLAS detector at the LHC. Events with three, four, or five or more reconstructed electrons and muons are analysed. The measured cross section for the $pp \rightarrow VVZ$ process is $660^{+93}_{-90}(\text{stat.})^{+88}_{-81}(\text{syst.})$ fb, and the observed (expected) significance is 6.4 (4.7) standard deviations, representing the first observation of $VVZ$ production. In addition, the measured cross section for the $pp \rightarrow WWZ$ process is $442 \pm 94(\text{stat.})^{+60}_{-52}(\text{syst.})$ fb, and the observed (expected) significance is 4.4 (3.6) standard deviations, representing evidence of $WWZ$ production. The measured cross sections are consistent with the SM predictions.
Constraints on physics beyond the SM are also derived in the EFT framework by setting limits on Wilson coefficients for dimension-8 operators describing anomalous quartic gauge boson couplings. The constraints are comparable to published limits in the $W\gamma jj$ final state.

\section*{Acknowledgements}
%

%
%

%
%

We thank CERN for the very successful operation of the LHC and its injectors, as well as the support staff at
CERN and at our institutions worldwide without whom ATLAS could not be operated efficiently.

The crucial computing support from all WLCG partners is acknowledged gratefully, in particular from CERN, the ATLAS Tier-1 facilities at TRIUMF/SFU (Canada), NDGF (Denmark, Norway, Sweden), CC-IN2P3 (France), KIT/GridKA (Germany), INFN-CNAF (Italy), NL-T1 (Netherlands), PIC (Spain), RAL (UK) and BNL (USA), the Tier-2 facilities worldwide and large non-WLCG resource providers. Major contributors of computing resources are listed in Ref.~\cite{ATL-SOFT-PUB-2025-001}.

We gratefully acknowledge the support of ANPCyT, Argentina; YerPhI, Armenia; ARC, Australia; BMWFW and FWF, Austria; ANAS, Azerbaijan; CNPq and FAPESP, Brazil; NSERC, NRC and CFI, Canada; CERN; ANID, Chile; CAS, MOST and NSFC, China; Minciencias, Colombia; MEYS CR, Czech Republic; DNRF and DNSRC, Denmark; IN2P3-CNRS and CEA-DRF/IRFU, France; SRNSFG, Georgia; BMBF, HGF and MPG, Germany; GSRI, Greece; RGC and Hong Kong SAR, China; ICHEP and Academy of Sciences and Humanities, Israel; INFN, Italy; MEXT and JSPS, Japan; CNRST, Morocco; NWO, Netherlands; RCN, Norway; MNiSW, Poland; FCT, Portugal; MNE/IFA, Romania; MSTDI, Serbia; MSSR, Slovakia; ARIS and MVZI, Slovenia; DSI/NRF, South Africa; MICIU/AEI, Spain; SRC and Wallenberg Foundation, Sweden; SERI, SNSF and Cantons of Bern and Geneva, Switzerland; NSTC, Taipei; TENMAK, T\"urkiye; STFC/UKRI, United Kingdom; DOE and NSF, United States of America.

Individual groups and members have received support from BCKDF, CANARIE, CRC and DRAC, Canada; CERN-CZ, FORTE and PRIMUS, Czech Republic; COST, ERC, ERDF, Horizon 2020, ICSC-NextGenerationEU and Marie Sk{\l}odowska-Curie Actions, European Union; Investissements d'Avenir Labex, Investissements d'Avenir Idex and ANR, France; DFG and AvH Foundation, Germany; Herakleitos, Thales and Aristeia programmes co-financed by EU-ESF and the Greek NSRF, Greece; BSF-NSF and MINERVA, Israel; NCN and NAWA, Poland; La Caixa Banking Foundation, CERCA Programme Generalitat de Catalunya and PROMETEO and GenT Programmes Generalitat Valenciana, Spain; G\"{o}ran Gustafssons Stiftelse, Sweden; The Royal Society and Leverhulme Trust, United Kingdom.

In addition, individual members wish to acknowledge support from Armenia: Yerevan Physics Institute (FAPERJ); CERN: European Organization for Nuclear Research (CERN DOCT); Chile: Agencia Nacional de Investigaci\'on y Desarrollo (FONDECYT 1230812, FONDECYT 1230987, FONDECYT 1240864); China: Chinese Ministry of Science and Technology (MOST-2023YFA1605700, MOST-2023YFA1609300), National Natural Science Foundation of China (NSFC - 12175119, NSFC 12275265, NSFC-12075060); Czech Republic: Czech Science Foundation (GACR - 24-11373S), Ministry of Education Youth and Sports (ERC-CZ-LL2327, FORTE CZ.02.01.01/00/22\_008/0004632), PRIMUS Research Programme (PRIMUS/21/SCI/017); EU: H2020 European Research Council (ERC - 101002463); European Union: European Research Council (ERC - 948254, ERC 101089007, ERC, BARD, 101116429), Horizon 2020 Framework Programme (MUCCA - CHIST-ERA-19-XAI-00), European Union, Future Artificial Intelligence Research (FAIR-NextGenerationEU PE00000013), Italian Center for High Performance Computing, Big Data and Quantum Computing (ICSC, NextGenerationEU); France: Agence Nationale de la Recherche (ANR-20-CE31-0013, ANR-21-CE31-0013, ANR-21-CE31-0022, ANR-22-EDIR-0002); Germany: Baden-Württemberg Stiftung (BW Stiftung-Postdoc Eliteprogramme), Deutsche Forschungsgemeinschaft (DFG - 469666862, DFG - CR 312/5-2); Italy: Istituto Nazionale di Fisica Nucleare (ICSC, NextGenerationEU), Ministero dell'Università e della Ricerca (NextGenEU PRIN20223N7F8K M4C2.1.1); Japan: Japan Society for the Promotion of Science (JSPS KAKENHI JP22H01227, JSPS KAKENHI JP22H04944, JSPS KAKENHI JP22KK0227, JSPS KAKENHI JP23KK0245); Norway: Research Council of Norway (RCN-314472); Poland: Ministry of Science and Higher Education (IDUB AGH, POB8, D4 no 9722), Polish National Agency for Academic Exchange (PPN/PPO/2020/1/00002/U/00001), Polish National Science Centre (NCN 2021/42/E/ST2/00350, NCN OPUS 2023/51/B/ST2/02507, NCN OPUS nr 2022/47/B/ST2/03059, NCN UMO-2019/34/E/ST2/00393, UMO-2020/37/B/ST2/01043, UMO-2021/40/C/ST2/00187, UMO-2022/47/O/ST2/00148, UMO-2023/49/B/ST2/04085, UMO-2023/51/B/ST2/00920); Portugal: Foundation for Science and Technology (FCT); Spain: Generalitat Valenciana (Artemisa, FEDER, IDIFEDER/2018/048), Ministry of Science and Innovation (MCIN \& NextGenEU PCI2022-135018-2, MICIN \& FEDER PID2021-125273NB, RYC2019-028510-I, RYC2020-030254-I, RYC2021-031273-I, RYC2022-038164-I); Sweden: Carl Trygger Foundation (Carl Trygger Foundation CTS 22:2312), Swedish Research Council (Swedish Research Council 2023-04654, VR 2018-00482, VR 2021-03651, VR 2022-03845, VR 2022-04683, VR 2023-03403), Knut and Alice Wallenberg Foundation (KAW 2018.0458, KAW 2019.0447, KAW 2022.0358); Switzerland: Swiss National Science Foundation (SNSF - PCEFP2\_194658); United Kingdom: Leverhulme Trust (Leverhulme Trust RPG-2020-004), Royal Society (NIF-R1-231091); United States of America: U.S. Department of Energy (ECA DE-AC02-76SF00515), Neubauer Family Foundation.

%
%


%
%
%

\printbibliography
\clearpage
 
\begin{flushleft}
\hypersetup{urlcolor=black}
{\Large The ATLAS Collaboration}

\bigskip

\AtlasOrcid[0000-0002-6665-4934]{G.~Aad}$^\textrm{\scriptsize 105}$,
\AtlasOrcid[0000-0001-7616-1554]{E.~Aakvaag}$^\textrm{\scriptsize 17}$,
\AtlasOrcid[0000-0002-5888-2734]{B.~Abbott}$^\textrm{\scriptsize 124}$,
\AtlasOrcid[0000-0002-0287-5869]{S.~Abdelhameed}$^\textrm{\scriptsize 120a}$,
\AtlasOrcid[0000-0002-1002-1652]{K.~Abeling}$^\textrm{\scriptsize 56}$,
\AtlasOrcid[0000-0001-5763-2760]{N.J.~Abicht}$^\textrm{\scriptsize 50}$,
\AtlasOrcid[0000-0002-8496-9294]{S.H.~Abidi}$^\textrm{\scriptsize 30}$,
\AtlasOrcid[0009-0003-6578-220X]{M.~Aboelela}$^\textrm{\scriptsize 46}$,
\AtlasOrcid[0000-0002-9987-2292]{A.~Aboulhorma}$^\textrm{\scriptsize 36e}$,
\AtlasOrcid[0000-0001-5329-6640]{H.~Abramowicz}$^\textrm{\scriptsize 158}$
% ...
\bigskip
\\

% ...
$^{17}$Department for Physics and Technology, University of Bergen, Bergen; Norway.\\
% ...
$^{30}$Physics Department, Brookhaven National Laboratory, Upton NY; United States of America.\\
% ...
$^{36}$$^{(a)}$Facult\'e des Sciences Ain Chock, Universit\'e Hassan II de Casablanca;$^{(b)}$Facult\'{e} des Sciences, Universit\'{e} Ibn-Tofail, K\'{e}nitra;$^{(c)}$Facult\'e des Sciences Semlalia, Universit\'e Cadi Ayyad, LPHEA-Marrakech;$^{(d)}$LPMR, Facult\'e des Sciences, Universit\'e Mohamed Premier, Oujda;$^{(e)}$Facult\'e des sciences, Universit\'e Mohammed V, Rabat;$^{(f)}$Institute of Applied Physics, Mohammed VI Polytechnic University, Ben Guerir; Morocco.\\
% ...
$^{46}$Physics Department, Southern Methodist University, Dallas TX; United States of America.\\
% ...
$^{50}$Fakult\"{a}t Physik , Technische Universit{\"a}t Dortmund, Dortmund; Germany.\\
% ...
$^{56}$II. Physikalisches Institut, Georg-August-Universit\"{a}t G\"ottingen, G\"ottingen; Germany.\\
% ...
$^{105}$CPPM, Aix-Marseille Universit\'e, CNRS/IN2P3, Marseille; France.\\
% ...
$^{120}$$^{(a)}$New York University Abu Dhabi, Abu Dhabi;$^{(b)}$United Arab Emirates University, Al Ain; United Arab Emirates.\\
% ...
$^{124}$Homer L. Dodge Department of Physics and Astronomy, University of Oklahoma, Norman OK; United States of America.\\
% ...
$^{158}$Raymond and Beverly Sackler School of Physics and Astronomy, Tel Aviv University, Tel Aviv; Israel.\\
% ...
$^{a}$ Also Affiliated with an institute covered by a cooperation agreement with CERN.\\
$^{b}$ Also at An-Najah National University, Nablus; Palestine.\\
$^{c}$ Also at Borough of Manhattan Community College, City University of New York, New York NY; United States of America.\\
$^{d}$ Also at Center for High Energy Physics, Peking University; China.\\
$^{e}$ Also at Center for Interdisciplinary Research and Innovation (CIRI-AUTH), Thessaloniki; Greece.\\
$^{f}$ Also at CERN, Geneva; Switzerland.\\
% ...

\end{flushleft}


%

%
%
\clearpage

\end{document}